\documentclass[aps,reprint,superscriptaddress,showpacs,floatfix]{revtex4-2}

\usepackage{feynmp}
\usepackage{amsmath}
\usepackage{graphicx}
\usepackage{multirow}
\usepackage{latexsym}
\usepackage{textcomp}
\usepackage{verbatim}
\usepackage{color}
\usepackage{bm}
\usepackage{subfigure}
\usepackage{everysel}
\usepackage{keyval}
\usepackage{ragged2e}
\usepackage{dsfont}
\usepackage{amssymb}
\usepackage{enumitem}
\usepackage{pstricks}
\usepackage{mathtools}
\usepackage{braket}
\usepackage{slashed} 
\usepackage{hyperref}
\hypersetup{
     colorlinks=true,
     allcolors = blue
     }

\usepackage{graphicx}
\usepackage{xcolor}

\usepackage{amsmath}
\usepackage{bbold}
\usepackage{amsfonts}
\usepackage{bbm}
\usepackage{orcidlink}

\newcommand{\osum}{{%
    \setbox0\hbox{\circ}%
    \rlap{\hbox to \wd0{\hss\sum\hss}}\box0
}}

\begin{document}

\title{Aharonov-Casher-induced electric quadrupole of charge-neutral particles}

\author{Hojun Lee\,\orcidlink{0000-0002-7406-1936}}
\thanks{These authors contributed equally to this work.}
\affiliation{Department of Physics, Pohang University of Science and Technology, Pohang 37673, Republic of Korea}
\affiliation{Center for Quantum Dynamics of Angular Momentum, Pohang 37673, Republic of Korea}

\author{Youngjae Jeon\,\orcidlink{0009-0009-4247-0899}}
\thanks{These authors contributed equally to this work.}
\affiliation{Department of Physics, Pohang University of Science and Technology, Pohang 37673, Republic of Korea}
\affiliation{Center for Quantum Dynamics of Angular Momentum, Pohang 37673, Republic of Korea}

\author{Suik Cheon\,\orcidlink{0000-0003-1789-3274}}
\email{enprodigy@postech.ac.kr}
\affiliation{Department of Physics, Pohang University of Science and Technology, Pohang 37673, Republic of Korea}
\affiliation{Center for Quantum Dynamics of Angular Momentum, Pohang 37673, Republic of Korea}

\author{Hyun-Woo Lee\,\orcidlink{0000-0002-1648-8093}}
\email{hwl@postech.ac.kr}
\affiliation{Department of Physics, Pohang University of Science and Technology, Pohang 37673, Republic of Korea}
\affiliation{Center for Quantum Dynamics of Angular Momentum, Pohang 37673, Republic of Korea}

\begin{abstract}
For charged particles, their orbital angular momentum (OAM) in solids has a direct magnetic manifestation as an orbital magnetization. For charge-neutral particles, however, the physical manifestation of the OAM in solids remains unclear. Here, we show that a charge-neutral particle carrying a magnetic moment couples to an electric-field gradient through the Aharonov–Casher (AC) effect, thereby exhibiting the electric quadrupole in crystalline solids. This AC-induced electric quadrupole (AC-EQ) contains the scalar, toroidal dipole, and reduced quadrupole components, which are conjugate to the divergence, circulation, and shear of the electric field, respectively. As representative examples, we calculate the AC-EQ of magnons in two magnetic systems. In ferromagnetic pyrochlore, Dzyaloshinskii–Moriya interaction (DMI) induces a sizable AC-EQ, whereas in helical $\mathrm{Fe}$ langasite the AC-EQ emerges from the helical spin configuration even in the absence of DMI.
\end{abstract}

\maketitle


\section{Introduction}\label{sec:Introduction}

The orbital motion of electrons in solids has long been a subject of great interest, as it constitutes, together with electron spin, the source of magnetism, one of the most important properties of materials. Unlike spin, which is an intrinsic degree of freedom of the electron, orbital motion is, however, subtle to formulate in a periodic crystal, where the position operator is ill-defined in the usual sense~\cite{Resta98PRL}. The modern theory of orbital magnetization resolves this difficulty by employing the Berry-phase formulation, where the position operator can be handled in the Bloch representation through derivatives with respect to crystal momentum~\cite{Xiao10RMP,Xiao05PRL,Thonhauser05PRL,Ceresoli06PRb,Shi07PRL}. This formalism successfully captures all the orbital contributions to magnetization, including both the local and itinerant circulation of electrons, and has enabled quantitative reproduction of orbital magnetization in crystalline solids~\cite{Ceresoli10PRb,Lopez12PRb,Hanke16PRb}.

The recent emergence of orbitronics, which aims to exploit the orbital degree of freedom of electrons as a carrier of information in quantum transport and information processing, has further amplified interest in the orbital angular momentum (OAM) of electrons~\cite{Bernevig05PRL,Kontani09PRL,Go18PRL,Go21EPL,Choi23NAT,Jo24NPJS,Wang25AElecM}. In particular, subsequent efforts have aimed to formulate electronic orbital responses, such as the orbital Hall effect, and the orbital Rashba-Edelstein effect, from the perspective of the modern theory of orbital magnetism~\cite{Pezo22PRB,Busch23PRR,Gobel24PRL,Lee24PRb,Liu25PRL,Cysne26arXiv,Sastges26arXiv}. These developments have opened a new route to understanding magnetic and transport phenomena in terms of orbital degrees of freedom, and more broadly highlight the importance of geometric quantities, such as Berry curvature and the quantum metric, in characterizing material properties~\cite{Lee25PRb,Lee26PRb}.

Recent studies have attempted to extend the modern theory of orbital magnetism of electrons to the orbital motion of bosonic quasiparticles, especially magnons. Neumann \textit{et al}. defined the orbital magnetic moment of magnons by evaluating the magnetic-field derivative of the thermodynamic energy and then separating the spin magnetic moment contribution~\cite{Neumann20PRL}. In a different approach, Fishman \textit{et al}. and subsequent studies evaluated the magnon OAM from the expectation value of the OAM operator~\cite{Fishman22PRL,Fishman23PRb,Fishman23JPc,Go24NanoLett}. A model calculation for a kagome antiferromagnet has shown that the magnon orbital magnetic moment and the magnon OAM are quantitatively distinct quantities~\cite{Jeon26arXiv}. This distinction reflects a fundamental difference between electrons and magnons: whereas the OAM of a charged electron is directly tied to its orbital magnetic moment and hence coupled to a magnetic field, the OAM of a charge-neutral magnon is not directly coupled to a magnetic field.

Instead, the orbital motion of charge-neutral particles carrying magnetic moments can couple to an electric field through the Aharonov-Casher (AC) effect~\cite{Aharonov84PRL,Cimmino89PRL,Mignani91JPa,Boyer97PRa,Kuzmenko25PRb}. At the Hamiltonian level, the AC effect can be formulated in terms of the AC coupling, in which the electric field and the magnetic moment enter as effective gauge potentials for the motion of the particle~\cite{Bakke09PRa,Nakata17PRb,Liu19PRb,Basso20EPL}. The AC coupling thereby provides a route for a charge-neutral particle carrying a magnetic moment to couple to an electric field, even in the absence of electric charge. Focusing on this coupling, Tang and Cheng recently formulated a proper theory of the magnon OAM as the thermodynamic response conjugate to the divergence of an electric field through the AC effect~\cite{Tang26PRL}, providing a basis for analyzing transport phenomena such as the magnon orbital Nernst effect. However, this coupling goes beyond just providing a rigorous definition of the magnon orbital angular momentum. It also implies that the orbital motion of charge-neutral particles can generate an equilibrium electric-quadrupole density. This electric response has been investigated in free space~\cite{Bakke19EPJP}, but there is no systematic formulation for it in crystalline solids.

Here, we develop a theory of the AC-induced electric quadrupole (AC-EQ) generated by the angular motion of charge-neutral particles carrying magnetic moments. The AC-EQ is defined as a thermodynamic response to the spatial variation of an electric field, in close analogy with the orbital magnetization of charged particles, which is defined as a thermodynamic response to a magnetic field. However, unlike the scalar electric charge, the magnetic moment is a vector quantity. As a result, the AC-EQ is described by a second-rank Cartesian tensor and contains richer information than the orbital magnetization or OAM vector, including both symmetric and antisymmetric components. Since Ref.~\cite{Tang26PRL} was aimed at deriving a proper expression for the magnon orbital angular momentum itself, it considered a fixed direction of the magnetic moment. From the perspective of the AC-EQ tensor, this captures only part of the symmetric component of this tensor, while the remaining components cannot be obtained from the response to the electric-field divergence alone. In the present theory, we formulate the full AC-EQ tensor, thereby encompassing all the information encoded in the electric quadrupole response generated by the orbital motion of charge-neutral particles carrying magnetic moments through the AC effect.

The rest of the article is organized as follows. Section~\ref{sec:Theory} develops the theory of the AC-EQ. We begin in Sec.~\ref{subsec:intuitive_picture} with an intuitive picture, showing that the angular motion of a charge-neutral point particle carrying a magnetic moment can generate the AC-EQ. Then, we derive the AC-EQ tensor in crystalline solids within a semiclassical framework in Sec.~\ref{subsec:semiclassical_derivation}. The irreducible decomposition is presented in Sec.~\ref{subsec:irreducible_decomposition}. Section~\ref{sec:Numerical_calculation} is devoted to numerical calculation of model systems. We apply our theory to magnons in two representative systems: the ferromagnetic pyrochlore $\mathrm{Lu_2V_2O_7}$ and the noncollinear $\mathrm{Fe}$ langasite $\mathrm{Ba_3NbFe_3Si_2O_{14}}$, which provide examples with and without DMI but with magnetic helicity, respectively. Finally, Sec.~\ref{sec:conclusion} summarizes and concludes the article with several remarks.

\section{Theoretical Formalism}\label{sec:Theory}

A standard way to describe the motion of a particle (or quasiparticle) in a solid is the semiclassical framework, which treats it as a wave packet and describes its dynamics. In this section, we show, based on the semiclassical approach, that the orbital angular motion of a charge-neutral particle carrying a magnetic moment can generate an electric quadrupole and derive the explicit form of the AC-EQ tensor. The angular motion of a wave packet contains two contributions: the self-rotation of the wave packet around its center and the center-of-mass motion of the wave-packet center. To build physical intuition for the emergence of an electric quadrupole, we first illustrate how orbital motion can generate an electric quadrupolar response, using a charge-neutral point particle carrying a magnetic moment in free space.

\subsection{Intuitive picture}\label{subsec:intuitive_picture}

\begin{figure}[t]
\includegraphics[width=245pt]{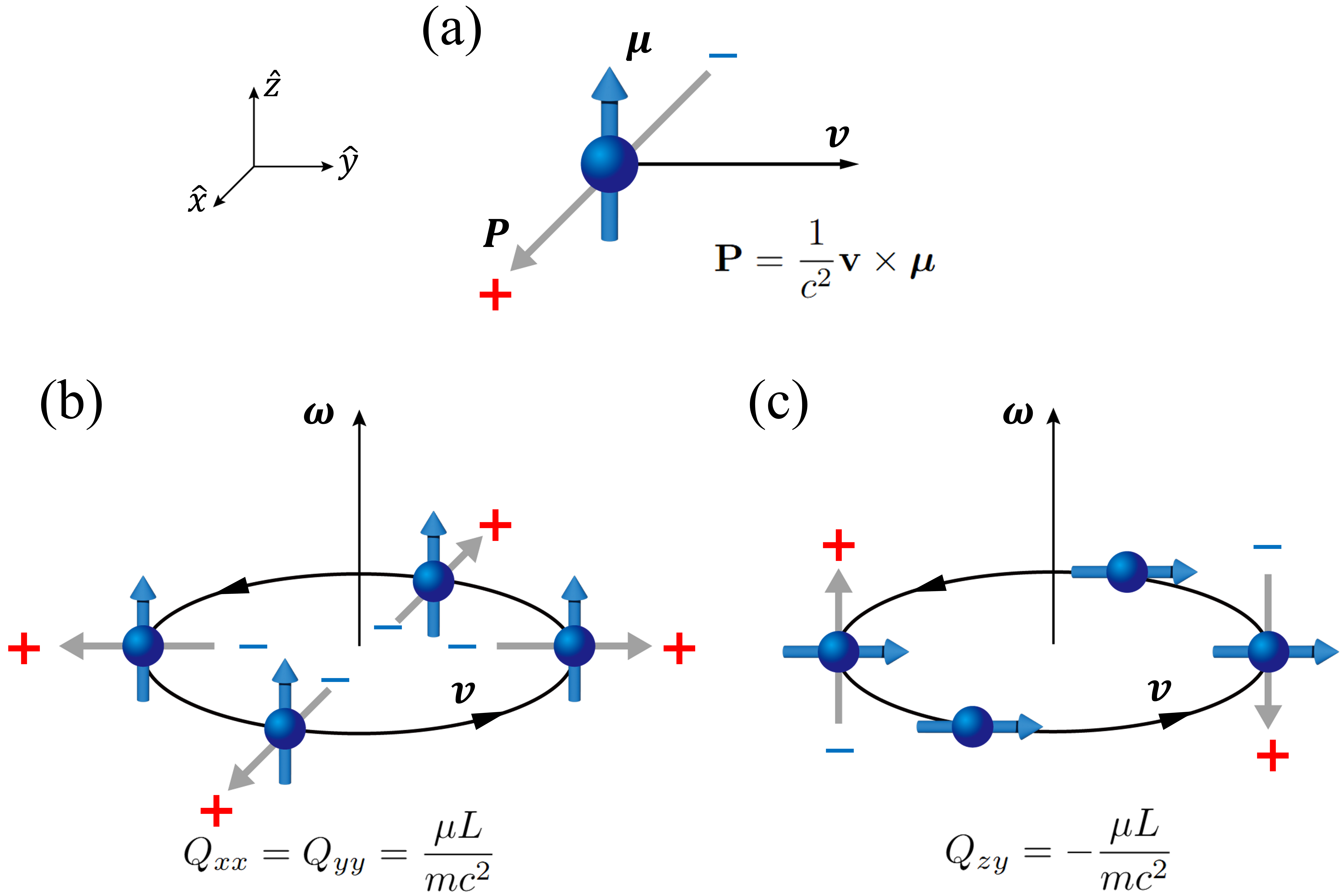}
\caption{\label{fig1}\textbf{Intuitive picture of Aharonov-Casher electric quadrupole (AC-EQ).} (a) The polarization $\mathbf{P}$ (gray arrow) induced by the linear motion of a particle carrying the magnetic moment $\bm{\mu}$. (b), (c) Representative local electric-quadrupolar contributions $Q_{\alpha \beta} = P_{\alpha}(\mathbf{u})u_{\beta}$ generated by the angular motion of a particle carrying $\bm{\mu}$. Considering the full orbit, the induced polarizations at opposite positions of the orbit cancel each other, whereas their spatial moments $P_{\alpha}u_{\beta}$ add constructively. (b) The case of $\bm{\mu} = \mu \hat{z}$ and $\bm{\omega} = \omega \hat{z}$. (c) The case of $\bm{\mu} = \mu \hat{y}$ and $\bm{\omega} = \omega \hat{z}$. The symbols + and - indicate local bound-charge distribution associated with the induced polarizations.}
\end{figure}

Consider a charge-neutral particle carrying a magnetic moment $\bm{\mu}$ and moving with velocity $\mathbf{v}$. Let $\mathbf{B}^{\prime}$ and $\mathbf{B}$ denote the magnetic fields measured in the rest frame of the particle $S^{\prime}$ and in the laboratory frame $S$, respectively. In the rest frame $S^{\prime}$, the Zeeman energy is given by
\begin{align}\label{eq:U_rest}
    U &= -\bm{\mu} \cdot \mathbf{B}^{\prime}
\end{align}
On the other hand, in the laboratory frame $S$, the energy is given by
\begin{align}\label{eq:U_lab}
    U &\approx -\bm{\mu} \cdot \left( \mathbf{B} - \frac{1}{c^{2}}\mathbf{v} \times \mathbf{E} \right)
\end{align}
in the limit $v \ll c$. Thus, the energy correction is given by
\begin{align}\label{eq:U_correction}
    \Delta U &= -\mathbf{P} \cdot \mathbf{E},
\end{align}
where
\begin{align}\label{eq:AC-polarization}
    \mathbf{P} &= \frac{1}{c^{2}} \mathbf{v} \times \bm{\mu}.
\end{align}
Here $\mathbf{P}$ can be interpreted as the motion-induced electric dipole moment of the particle carrying $\bm{\mu}$ [Fig.~\ref{fig1}(a)]. This relativistic motional-dipole coupling can be regarded as the AC coupling expressed in the energy form~\cite{Bakke09PRa}.

More generally, the motion of a charge-neutral particle carrying $\bm{\mu}$ can be described in terms of effective electromagnetic fields: the effective electric field
\begin{align}\label{eq:effective_E-field}
    \mathbf{E}_{\text{eff}} &= \frac{1}{\mu}\nabla (\bm{\mu} \cdot \mathbf{B})
\end{align}
through the Zeeman effect, which exerts the force $\mathbf{F} = \mu \mathbf{E}_{\text{eff}} = \nabla (\bm{\mu} \cdot \mathbf{B})$, and the effective magnetic field
\begin{align}\label{eq:effective_B-field}
    \mathbf{B}_{\text{eff}} &= \frac{1}{\mu c^{2}} \nabla \times (\bm{\mu} \times \mathbf{E})
\end{align}
through the AC effect, which exerts the force $\mathbf{F} = \mu \mathbf{v} \times \mathbf{B}_{\text{eff}}$. Here $\mu = \vert \bm{\mu} \vert$. The effective scalar and vector potentials can be chosen to be
\begin{align}\label{eq:effective_potentials}
    \phi_{\text{eff}} = -\frac{1}{\mu}(\bm{\mu} \cdot \mathbf{B}), \quad \mathbf{A}_{\text{eff}} = \frac{1}{\mu c^{2}} (\bm{\mu} \times \mathbf{E}),
\end{align}
such that $\mathbf{E}_{\text{eff}} = -\nabla \phi_{\text{eff}}$ and $\mathbf{B}_{\text{eff}} = \nabla \times \mathbf{A}_{\text{eff}}$.

We now consider the situation in which the particle undergoes circular motion around an axis at $\mathbf{r}_{\text{o}}$ with angular velocity $\bm{\omega}$. Then, the velocity is given by $\mathbf{v} = \bm{\omega} \times \mathbf{u}$, where $\mathbf{u} = \mathbf{r} - \mathbf{r}_{\text{o}}$ is the relative position of the particle. Substituting this into Eq.~\eqref{eq:AC-polarization}, we obtain
\begin{align}\label{eq:toy-polarization}
    \mathbf{P} &= \frac{1}{c^{2}}\left[ \mathbf{u}(\bm{\omega} \cdot \bm{\mu}) - \bm{\omega} (\mathbf{u} \cdot \bm{\mu})\right] .
\end{align}
The first term on the right-hand side of Eq.~\eqref{eq:toy-polarization} is the radial part, which is parallel to $\mathbf{u}$, and the second term is the axial part, which is parallel to $\bm{\omega}$.

The circular motion of a particle generates a spatial pattern of local electric dipoles rather than a uniform electric dipole. Consider two diametrically opposite points of the orbit. Under the transformation $\mathbf{u} \rightarrow -\mathbf{u}$, the velocity also changes sign, $\mathbf{v} \rightarrow -\mathbf{v}$. For a fixed magnetic moment $\bm{\mu}$, Eq.~\eqref{eq:AC-polarization} gives $\mathbf{P} \rightarrow -\mathbf{P}$. Thus, the dipolar contributions from opposite points cancel when the full orbit is considered. On the other hand, the spatial moment $P_{\alpha}u_{\beta}$ remains invariant, and these quadrupolar contributions add constructively around the orbit. This provides the simplest physical picture of how orbital motion can have no net electric dipole but generate a finite electric-quadrupole response.

To make this explicit, we consider a spatially varying electric field. The interaction energy at the instantaneous particle position $\mathbf{r} = \mathbf{r}_{\text{o}} + \mathbf{u}$ is given by
\begin{align}
    \Delta U (\mathbf{r}_{\text{o}}) &= -P_{\alpha}(\mathbf{u}) E_{\alpha}(\mathbf{r}_{\text{o}} + \mathbf{u}),
\end{align}
where
\begin{align}
    E_{\alpha}(\mathbf{r}_{\text{o}}+\mathbf{u}) &= E_{\alpha}(\mathbf{r}_{\text{o}}) + u_{\beta}\partial_{\beta}E_{\alpha}(\mathbf{r}_{\text{o}}) + \cdots .
\end{align}
The uniform-field term couples to the local electric dipole and cancels between opposite positions of the closed orbit, as discussed above. Then, the leading term is proportional to the electric-field gradient, 
\begin{align}\label{eq:E-expanding}
    \Delta U &\approx - P_{\alpha} E_{\alpha} - Q_{\alpha \beta} \partial_{\beta} E_{\alpha},
\end{align}
where
\begin{align}\label{eq:AC-EQ_simple}
    Q_{\alpha \beta} &= P_{\alpha}u_{\beta} = \frac{1}{mc^{2}} \left[ (\bm{\omega} \cdot \bm{\mu})I_{\alpha \beta} - \omega_{\alpha} \mu_{\gamma}I_{\beta \gamma}\right]
\end{align}
is the local electric quadrupolar contribution associated with a given position along the orbit. Here, $I_{\alpha \beta} = m u_{\alpha} u_{\beta}$ is the second moment tensor, and $\alpha, \beta, \gamma \in \{x, y, z\}$. The quadrupolar response of the orbit motion arises from the spatial pattern of these contributions, which add rather than cancel at diametrically opposite points. The corresponding bulk thermodynamic AC-EQ in crystalline solid is formulated systematically in Sec.~\ref{subsec:semiclassical_derivation}.

We emphasize that this second-rank tensor is not restricted to the symmetric traceless electric quadrupole, which we refer to as the reduced electric quadrupole, but also contains rank-0 and rank-1 components. The corresponding irreducible decomposition is discussed in Sec.~\ref{subsec:irreducible_decomposition}.

Now, we illustrate this local quadrupolar pattern for two representative cases. First, when $\bm{\mu}$ is parallel to $\bm{\omega}$ ($\bm{\mu} = \mu \hat{z}$ and $\bm{\omega} = \omega \hat{z}$) [Fig.~\ref{fig1}(b)], the axial part of $\mathbf{P}$ vanishes, and $\mathbf{P}$ points radially in the orbital plane. This gives rise to diagonal components such as $Q_{xx} = \mu u_{x}^{2}\omega / c^{2}$ and $Q_{yy} = \mu u_{y}^{2}\omega / c^{2}$. For a circular orbit of radius $u$, their representative local values at the points on the $x$ and $y$ axes are $Q_{xx} = Q_{yy} = \mu L / m c^{2}$, where $L = m u^{2} \omega$ is the orbital angular momentum. As presented in Fig.~\ref{fig1}(b), the dipoles at opposite positions point in opposite directions, whereas the associated $Q_{xx}$ or $Q_{yy}$ contributions have the same sign.

Next, when $\bm{\mu}$ is perpendicular to $\bm{\omega}$ ($\bm{\mu} = \mu \hat{y}$ and $\bm{\omega} = \omega \hat{z}$) [Fig.~\ref{fig1}(c)], the radial part of $\mathbf{P}$ vanishes, and $\mathbf{P}$ points along the rotation axis. At the representative points on the $y$ axis, this gives the off-diagonal local contribution $Q_{zy} = -\mu L / mc^{2}$. Again, the dipolar contributions cancel between opposite points, whereas the corresponding $Q_{zy}$ contributions add.

\subsection{Semiclassical derivation}\label{subsec:semiclassical_derivation}

We now derive the AC-EQ within the semiclassical wave-packet formalism. Unlike a single point particle carrying a fixed magnetic moment $\bm{\mu}$, a wave packet in a crystal generally carries a band-projected magnetic moment $\bm{\mu}_{n\mathbf{k}}$ with a nontrivial momentum-space texture, where $n$ denotes the band index and $\mathbf{k}$ the crystal momentum. Therefore, it is convenient to absorb the normalization factor $1/\mu$ in Eqs.~\eqref{eq:effective_E-field}-\eqref{eq:effective_potentials}, which is appropriate for a fixed-magnitude magnetic moment, into the effective electromagnetic fields and potentials. We thus introduce
\begin{align}\label{eq:effective_fields_tilde}
    \widetilde{\mathbf{E}}_{\text{eff}} = \nabla (\bm{\mu} \cdot \mathbf{B}), \quad \widetilde{\mathbf{B}}_{\text{eff}} = \frac{1}{c^{2}} \nabla \times (\bm{\mu} \times \mathbf{E}),
\end{align}
and the corresponding effective potentials
\begin{align}\label{eq:effective_potentials_tilde}
    \widetilde{\phi}_{\text{eff}} = -(\bm{\mu} \cdot \mathbf{B}), \quad \widetilde{\mathbf{A}}_{\text{eff}} = \frac{1}{c^{2}} (\bm{\mu} \times \mathbf{E}).
\end{align}
Here, $\bm{\mu} = \bm{\mu}_{n\mathbf{k}}$ denotes the band-projected magnetic dipole moment that enters the AC coupling of the wave packet. We note that the effective vector potential $\widetilde{\mathbf{A}}_{\text{eff}}$ is directly related to the conventional AC phase $\phi_{\text{AC}}$ as
\begin{align}
    \phi_{\text{AC}} &= \frac{1}{\hbar}\oint_{\mathcal{C}}\widetilde{A}_{\text{eff}}\cdot d\mathbf{r} = \frac{1}{\hbar} \int \widetilde{\mathbf{B}}_{\text{eff}} \cdot d\mathbf{s}.
\end{align}
Let $\hat{H}_{0}$ be the unperturbed bare Bloch Hamiltonian. Then, the local Hamiltonian for a wave packet centered at $\mathbf{r}_{\text{c}}$ is written as
\begin{align}\label{eq:local_Hamiltonian_WP}
    \hat{H}_{\textrm{c}} &= \hat{H}_{0}\left( \textbf{q} - \frac{1}{\hbar}\widetilde{\textbf{A}}_{\textrm{eff}} (\textbf{r}_{\textrm{c}}) \right) + \widetilde{\phi}_{\textrm{eff}}(\textbf{r}_{\textrm{c}}),
\end{align}
where $\mathbf{q}$ is the canonical wave vector and $\mathbf{k} = \mathbf{q} - \widetilde{\mathbf{A}}_{\text{eff}}/\hbar$ is the kinematic wave vector. The spatial variation of the effective magnetic field $\widetilde{\mathbf{B}}_{\text{eff}}(\mathbf{r})$ gives the gradient correction to the energy
\begin{align}\label{eq:energy_correction}
    \Delta \mathcal{E} &= -\textbf{m}_{\textrm{E}} \cdot \widetilde{\textbf{B}}_{\textrm{eff}},
\end{align}
where
\begin{align}\label{eq:small_me}
    \textbf{m}_{\textrm{E}, n\mathbf{k}} &= \frac{1}{2\hbar} \, \textrm{Im} \left[ \braket{\nabla_{\textbf{k}}u_{n\mathbf{k}} | \times (\mathcal{E}_{n\mathbf{k}}  - \hat{H}_{0}(\textbf{k})) | \nabla_{\textbf{k}}u_{n\mathbf{k}}} \right]
\end{align}
is the orbital AC moment of the Bloch states. It represents the self-rotation of the wave packet in the effective field~\cite{Sundaram99PRb}.

Meanwhile, the effective magnetic field modifies the volume element of the phase space $\Delta V = \Delta \mathbf{r} \Delta \mathbf{k}$. It forces the introduction of a modified density of states
\begin{align}\label{eq:modified_DOS}
    \mathcal{D}_{n\mathbf{k}} &= (2\pi)^{-d}(1 - \frac{1}{\hbar}\widetilde{\textbf{B}}_{\textrm{eff},n\mathbf{k}} \cdot \bm{\Omega}_{n\mathbf{k}}),
\end{align}
where
\begin{align}\label{eq:Berry_curvature}
    \bm{\Omega}_{n\mathbf{k}} &= -\textrm{Im} \left[ \braket{\nabla_{\textbf{k}}u_{n\mathbf{k}} | \times | \nabla_{\textbf{k}}u_{n\mathbf{k}}} \right]
\end{align}
is the Berry curvature~\cite{Xiao05PRL}.
Thus, the total energy of the system is given by
\begin{align}\label{eq:total_energy}
    \mathcal{E}_{\textrm{tot}} &= \sum_{n} \int d\textbf{k}\, f_{n\mathbf{k}}\mathcal{D}_{n\mathbf{k}}\left(\mathcal{E}_{n\mathbf{k}} - \textbf{m}_{\textrm{E},n\mathbf{k}} \cdot \widetilde{\textbf{B}}_{\textrm{eff},n\mathbf{k}}\right).
\end{align}

The thermodynamic response is obtained from the grand potential $G = \mathcal{E}_{\text{tot}} - \mu_{\text{ch}} \mathcal{N} - TS$. We define the AC-EQ tensor as the response conjugate to the electric-field gradient as follows:
\begin{align}\label{eq:AC-EQ_tensor}
    \mathcal{Q}_{\alpha \beta} &= -\frac{\partial G}{\partial (\partial_{\beta}E_{\alpha})}\notag
    \\[2mm]
    &= \frac{1}{c^{2}} \sum_{n} \int \frac{d\textbf{k}}{(2\pi)^{d}} \,\left[ (\boldsymbol{\mu}_{n\mathbf{k}} \cdot \boldsymbol{\Lambda}_{n\mathbf{k}})\delta_{\alpha \beta} - \mu_{n\mathbf{k},\beta}\Lambda_{n\mathbf{k},\alpha} \right],
\end{align}
where
\begin{align}\label{eq:Lambda}
    \Lambda_{n\mathbf{k}} &= F_{n\mathbf{k}} \mathbf{m}_{\textrm{E},n\mathbf{k}} \pm \frac{1}{\hbar \beta}\bm{\Omega}_{n\mathbf{k}}\ln{\left( 1 \mp e^{-\beta (\mathcal{E}_{n\mathbf{k}}-\mu_{\textrm{ch}})} \right)}.
\end{align}
Here, $F_{n\mathbf{k}} \equiv 1/(e^{\beta (\mathcal{E}_{n\mathbf{k}} - \mu_{\textrm{ch}})} \mp 1)$ is the equilibrium distribution function and $\mu_{\textrm{ch}}$ is the chemical potential. For both $\Lambda_{n\mathbf{k}}$ and $F_{n\mathbf{k}}$, the upper and lower signs correspond to Bosons and Fermions, respectively. The first term in $\bm{\Lambda}_{n\mathbf{k}}$ gives the self-rotation contribution, whereas the second term gives the center-of-mass motion contribution of the AC-EQ. Since Eq.~\eqref{eq:AC-EQ_tensor} gives a bulk thermodynamic response per unit volume, $\mathcal{Q}_{\alpha \beta}$ should be understood as the AC-EQ density. For convenience, we refer to this quantity simply as the AC-EQ unless the distinction between density and moment is important. In this paper, the AC-EQ denotes the full reducible second-rank electric multipole tensor conjugate to $\partial_{\beta}E_{\alpha}$, rather than only its symmetric traceless component, which we refer to as the reduced electric quadrupole in the following section. We note that the electric quadrupole of the point-particle picture [Eq.~\eqref{eq:AC-EQ_simple}] is promoted in solids to the band-geometric formula, with the classical orbital motion replaced by $\bm{\Lambda}_{n\mathbf{k}}$ [Eq.~\eqref{eq:AC-EQ_tensor}]. The details of the derivation are presented in Appendix~\ref{appendix:Derivation_semiclassical}.

\subsection{Irreducible decomposition}\label{subsec:irreducible_decomposition}

The AC-EQ tensor in Eq.~\eqref{eq:AC-EQ_tensor} is a second-rank Cartesian tensor. We introduce the irreducible decomposition of the AC-EQ tensor as follows:
\begin{align}\label{eq:irreducible_decomposition}
    \mathcal{Q}_{\alpha \beta} &= \mathcal{Q}^{(0)}_{\alpha \beta} + \mathcal{Q}^{(1)}_{\alpha \beta} + \mathcal{Q}^{(2)}_{\alpha \beta}.
\end{align}
The rank-0 component is the isotropic part,
\begin{align}\label{eq:scalar_AC-EQ}
    \mathcal{Q}^{(0)}_{\alpha \beta} &= \frac{1}{3} \delta_{\alpha \beta} \mathcal{Q}_{\gamma \gamma}\notag
    \\[2mm]
    &= \frac{2}{3c^{2}} \sum_{n} \int \frac{d\textbf{k}}{(2\pi)^{d}} \,(\boldsymbol{\mu}_{n\mathbf{k}} \cdot \boldsymbol{\Lambda}_{n\mathbf{k}})\delta_{\alpha \beta}.
\end{align}
We call this component the AC-electric monopole (AC-EM). It has one independent degree of freedom. We note that the AC-EM does not represent a net electric charge, but rather denotes the scalar trace component of the AC-EQ tensor. The rank-1 component is the antisymmetric part,
\begin{align}\label{eq:AC-ETD}
    \mathcal{Q}^{(1)}_{\alpha \beta} &= \frac{1}{2}(\mathcal{Q}_{\alpha \beta} - \mathcal{Q}_{\beta \alpha}) = \frac{1}{2}\varepsilon_{\alpha \beta \gamma}T^{\text{AC}}_{\gamma},
\end{align}
where
\begin{align}\label{eq:AC-ETD_vector}
    T^{\text{AC}}_{\gamma} &= \varepsilon_{\gamma \alpha \beta}\mathcal{Q}_{\alpha \beta}\notag
    \\[2mm]
    &= \frac{1}{c^{2}} \sum_{n} \int \frac{d\textbf{k}}{(2\pi)^{d}} \,(\boldsymbol{\mu}_{n\mathbf{k}} \times \boldsymbol{\Lambda}_{n\mathbf{k}}).
\end{align}
We call $\mathbf{T}^{\text{AC}}$ the Aharonov-Casher-induced electric toroidal dipole (AC-ETD). It has three independent components. A convenient basis for the AC-ETD is $\{T_{x}, T_{y}, T_{z}\}$. The rank-2 component is the symmetric traceless part,
\begin{align}
    \mathcal{Q}^{(2)}_{\alpha \beta} &= \frac{\mathcal{Q}_{\alpha \beta} + \mathcal{Q}_{\beta \alpha}}{2} - \frac{1}{3} \delta_{\alpha \beta} \mathcal{Q}_{\gamma \gamma}\notag
    \\[2mm]
    &= \frac{1}{c^{2}} \sum_{n} \int \frac{d\textbf{k}}{(2\pi)^{d}} \,\biggr[ \frac{1}{3}(\boldsymbol{\mu}_{n\mathbf{k}} \cdot \boldsymbol{\Lambda}_{n\mathbf{k}})\delta_{\alpha \beta}\notag
    \\[2mm]
    &\qquad \qquad - \frac{1}{2}(\mu_{n\mathbf{k},\alpha}\Lambda_{n\mathbf{k},\beta} + \mu_{n\mathbf{k},\beta}\Lambda_{n\mathbf{k},\alpha}) \biggr].
\end{align}
We call this component the Aharonov-Casher-induced reduced electric quadrupole (AC-REQ). It has five independent components. A common basis of the AC-REQ is given by $\{ \mathcal{Q}^{(2)}_{3z^{2}-r^{2}}, \mathcal{Q}^{(2)}_{x^{2}-y^{2}},  \mathcal{Q}^{(2)}_{xy}, \mathcal{Q}^{(2)}_{yz}, \mathcal{Q}^{(2)}_{zx} \}$.

\begin{figure}[t]
\includegraphics[width=245pt]{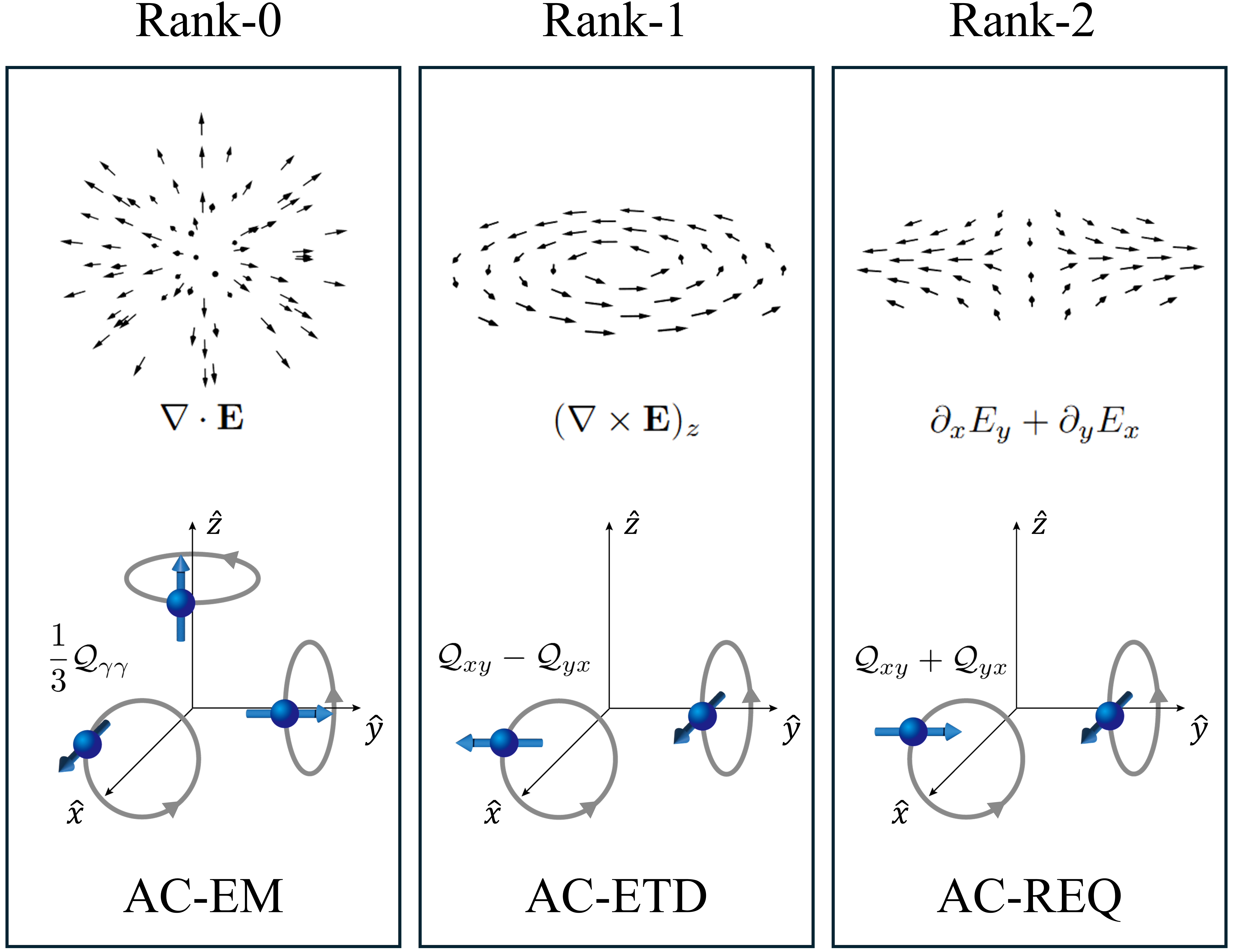}
\caption{\label{fig2}\textbf{Irreducible decomposition of AC-EQ.} The three columns represent the rank-0 AC-EM, rank-1 Aharonov-Casher electric toroidal dipole (AC-ETD), and rank-2 reduced AC-EQ (AC-REQ), respectively. The upper panels show the electric-field configurations conjugate to the respective AC-EQ components. The lower panels schematically illustrate the corresponding combinations of the magnetic moment $\bm{\mu}$ and the orbital motion of the particle.}
\end{figure}

The physical meaning of this decomposition becomes transparent by considering the electric-field-gradient tensor $E_{\alpha \beta} = \partial_{\beta}E_{\alpha}$. $E_{\alpha \beta}$ can be decomposed into its irreducible components as
\begin{align}\label{eq:appendix_E_irreducible_decomposition}
    E_{\alpha \beta} &= E_{\alpha \beta}^{(0)} + E_{\alpha \beta}^{(1)} + E_{\alpha \beta}^{(2)},
\end{align}
where
\begin{align}\label{eq:appendix_div_E}
    E_{\alpha \beta}^{(0)} &= \frac{1}{3}\delta_{\alpha \beta} (\nabla \cdot \mathbf{E})
\end{align}
is the divergence part,
\begin{align}\label{eq:E-tensor_divergence}
    E_{\alpha \beta}^{(1)} &= \frac{1}{2}(\partial_{\beta}E_{\alpha} - \partial_{\alpha}E_{\beta}) = \frac{1}{2}\varepsilon_{\gamma \beta \alpha}(\nabla \times \mathbf{E})_{\gamma}
\end{align}
is the circulation part, and
\begin{align}\label{eq:E-tensor_circulation}
    E_{\alpha \beta}^{(2)} &= \frac{1}{2}(\partial_{\beta}E_{\alpha} + \partial_{\alpha}E_{\beta}) - \frac{1}{3}\delta_{\alpha \beta} (\nabla \cdot \mathbf{E})
\end{align}
is the shear part. From the chain rule, the derivatives with respect to the divergence and circulation fields are given by
\begin{align}\label{eq:appendix_derivative_E-tensor_divergence}
    \frac{\partial G}{\partial (\nabla \cdot \mathbf{E})} &= \frac{1}{3}\delta_{\alpha \beta}\frac{\partial G}{\partial (\partial_{\beta}E_{\alpha})},
\end{align}
and
\begin{align}\label{eq:appendix_derivative_E-tensor_circulation}
    \frac{\partial G}{\partial (\nabla \times \mathbf{E})_{\gamma}} &= \frac{1}{2}\varepsilon_{\gamma \beta \alpha}\frac{\partial G}{\partial (\partial_{\beta}E_{\alpha})}.
\end{align}
These relations show that the AC-EM and the AC-ETD are the responses conjugate to the divergence and circulation parts of $E_{\alpha \beta}$, respectively. More generally, the variation of $G$ can be expressed as
\begin{align}\label{eq:appendix_variation_G}
    dG &= -\mathcal{Q}_{\alpha \beta} \, dE_{\alpha \beta}.
\end{align}
Substituting the irreducible decompositions of both $\mathcal{Q}_{\alpha \beta}$ and $E_{\alpha \beta}$, and the orthogonality between different irreducible sectors, we obtain
\begin{align}\label{eq:appendix_variation_G_decomposition}
    dG &= -\mathcal{Q}_{\alpha \beta}^{(0)}dE_{\alpha \beta}^{(0)} -\mathcal{Q}_{\alpha \beta}^{(1)}dE_{\alpha \beta}^{(1)} -\mathcal{Q}_{\alpha \beta}^{(2)}dE_{\alpha \beta}^{(2)}.
\end{align}
This implies that each irreducible component of $\mathcal{Q}_{\alpha \beta}$ is conjugate to the corresponding irreducible component of $E_{\alpha \beta}$. Thus, the AC-EM, AC-ETD, and AC-REQ are associated with the divergence part, circulation part, and shear part, respectively.

The AC-EQ is even under both spatial inversion and time reversal. Consequently, neither inversion nor time-reversal symmetry forbids the existence of the AC-EQ. Under spatial symmetries, the three irreducible parts of the AC-EQ tensor transform differently. The AC-EM is invariant under any point-group operation, the AC-ETD transforms as an axial vector, and the AC-REQ transforms as a symmetric traceless rank-two tensor. Thus, the point group determines which components of the AC-ETD and AC-REQ can remain finite.

For electrostatic fields, $E_{\alpha \beta}^{(1)}$, proportional to $\nabla \times \mathbf{E}$, vanishes by Faraday's law. Thus, the AC-ETD should be understood either as a response to a generalized electric-field-gradient perturbation or as a component accessible in time-dependent field configurations, whereas the AC-EM and AC-REQ are directly relevant to electrostatic field gradients.

We close this section by noting that the present formulation applies generally to any particle or quasiparticle carrying magnetic moments, whose motion can give rise to AC-induced electric multipole responses. In the following section, we explore the AC-EQ further by focusing on magnons as a representative system.

\section{Numerical Calculation}\label{sec:Numerical_calculation}

\begin{figure}[t]
\includegraphics[width=\linewidth]{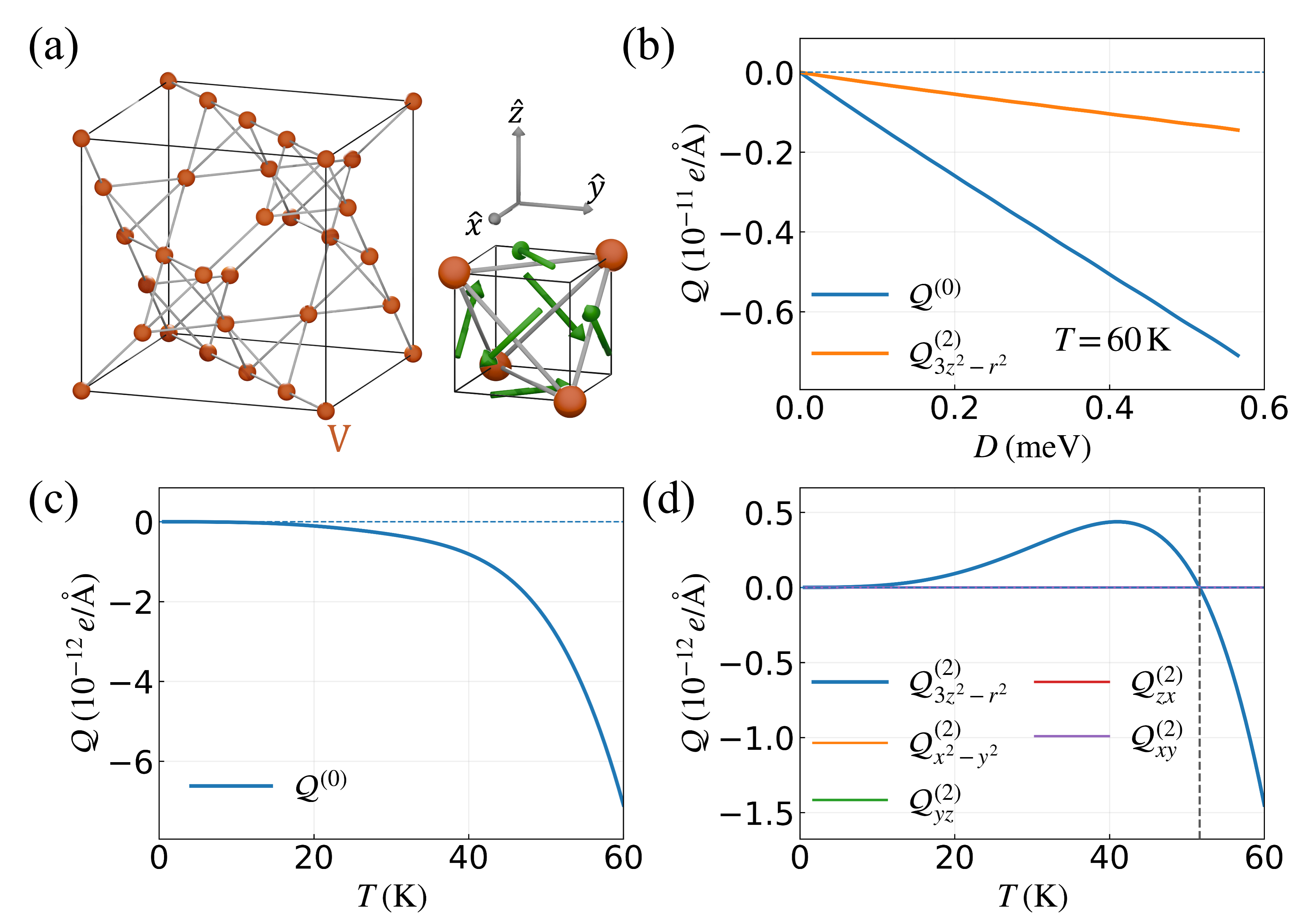}
\caption{\label{fig3} Crystal structure, DMI dependence, and temperature dependence of the AC-EQ components in ferromagnetic pyrochlore $\mathrm{Lu_2V_2O_7}$. (a) Left: crystal structure of $\mathrm{Lu_2V_2O_7}$ in the conventional unit cell. Right: four-sublattice basis, where the green arrows indicate the directions of the DMI vectors. Only $\mathrm{V}$ atoms are shown as spheres. (b) DMI dependence of the AC-EM and $\mathcal{Q}^{(2)}_{3z^2-r^2}$, calculated at fixed temperature $T=60$ K. (c), (d) Temperature dependence of the AC-EM and AC-REQ components, respectively. These results are calculated for $|\bm{D}|\approx 0.57\,\mathrm{meV}$.}
\end{figure}
In this section, we employ magnons as a platform for demonstrating the AC-EQ. Although any particle or quasiparticle with a magnetic moment can in principle exhibit the AC-EQ, its relativistic origin may make it challenging to isolate in charged systems, where ordinary electric coupling and screening effects can dominate. As charge-neutral quasiparticles carry magnetic moments and can be controlled and detected experimentally, magnons serve as a natural platform for this purpose.

To evaluate the AC-EQ tensor in Eq.~\eqref{eq:AC-EQ_tensor}, we take $\bm{\mu}_{n\mathbf{k}}$ to be the equilibrium magnetic moment entering the AC coupling. From a thermodynamic viewpoint, the magnetic moment associated with an individual magnon mode is obtained from the magnetic-field derivative of its band energy, and the total thermodynamic magnetization is obtained from the derivative of the full free energy~\cite{Neumann20PRL}. It can be decomposed into the spin-Zeeman contribution, often referred to as the magnon spin magnetic moment, and the remaining non-spin contribution, often referred to as the magnon orbital magnetic moment. Since magnons are charge neutral, the latter should not be understood as a direct coupling between the orbital motion of the magnon and a magnetic field, but rather as the part of the magnetic-field response not directly attributable to the spin Zeeman term. $\Lambda_{n\mathbf{k}}$ is determined by the magnon band structure and describes the orbital motion of the magnon wave packet, including self-rotation and center-of-mass motion contributions. In this work, we set the magnon chemical potential $\mu_{\text{ch}}$ to zero, as appropriate for equilibrium magnons whose number is not conserved. A finite $\mu_{\text{ch}}$ may arise in driven or pumped magnon systems, but such nonequilibrium situations are beyond the scope of the present work. We note that the expression in Eq.~\eqref{eq:AC-EQ_tensor} contains the contribution from thermally populated magnons. As shown in Appendix~\ref{Appendix:ZP}, the explicit harmonic zero-point contribution does not generate an additional AC-EQ. Calculation details, including the model parameters, numerical implementation, and magnonic band structure, are provided in the Supplemental Material~\cite{suppl_ref}.

First, we calculate the AC-EQ of the ferromagnetic pyrochlore $\mathrm{Lu_2V_2O_7}$, which possesses a sizable Dzyaloshinskii–Moriya interaction (DMI). We consider the ferromagnetic ground state with spins aligned with the $z$ axis in Fig.~\ref{fig3}(a). As a relativistic interaction in the system, DMI gives rise to a nontrivial magnon Berry curvature~\cite{Onose10SCI,Mena14PRL} and is likewise expected to generate a nontrivial magnon OAM texture. Moreover, a nontrivial magnon orbital magnetic moment texture has been theoretically reported in this material~\cite{Neumann20PRL}. The DMI-induced OAM texture and orbital magnetic moment texture in $\mathrm{Lu_2V_2O_7}$ make this material a promising candidate for exhibiting finite AC-EQ components. The effective spin Hamiltonian is given by
\begin{align}\label{FM_pyrochlore_spin_Hamiltonian}
    \hat{H}=\sum_{\left<i,j\right>}\left(-J\,\hat{\mathbf{S}}_i\cdot\hat{\mathbf{S}}_j + \bm{D}_{ij}\cdot\hat{\mathbf{S}}_i\times\hat{\mathbf{S}}_j \right)-g\mu_B\mathbf{B}\cdot\sum_{i}\hat{\mathbf{S}}_i,
\end{align}
where $J$ denotes ferromagnetic exchange coupling, $\hat{\mathbf{S}}_i$ is spin at site $i$, and $\bm{D}_{ij}$ denotes the DMI vector associated with bond $ij$. The corresponding DMI vectors are indicated with green arrows in Fig.~\ref{fig3}(a).

\begin{figure}[t]
\includegraphics[width=\linewidth]{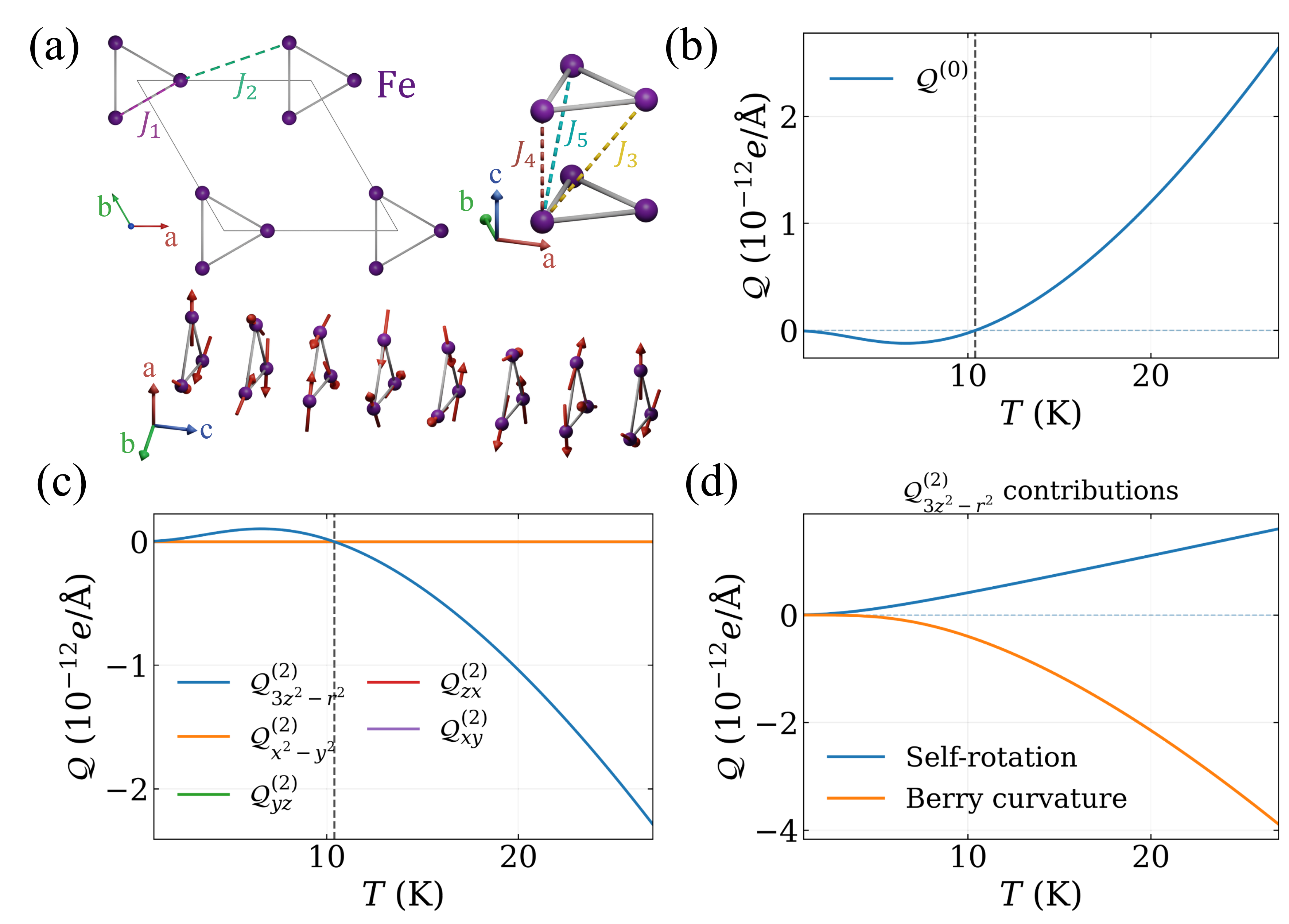}
\caption{\label{fig4} Crystal structure, magnetic structure, and temperature dependence of the AC-EQ components in noncollinear antiferromagnetic $\mathrm{Fe}$ langasite $\mathrm{Ba_3NbFe_3Si_2O_{14}}$. (a) Top left: top view of the crystal structure of $\mathrm{Fe}$ langasite, where only the $\mathrm{Fe}$ atoms are shown. $J_1$ and $J_2$ denote intralayer exchange paths. Top right: interlayer exchange paths $J_3$, $J_4$, and $J_5$. Bottom: helical magnetic structure of $\mathrm{Fe}$ langasite. We adopt a commensurate approximation in which seven magnetic layers constitute one magnetic unit cell. The crystallographic $\mathrm{a}$ and $\mathrm{c}$ axes are chosen as the $x$ and $z$ axes, respectively. (b), (c) Temperature dependence of the AC-EM and AC-REQ components, respectively. (d) Self-rotation and Berry curvature contributions to $\mathcal{Q}^{(2)}_{3z^2-r^2}$.}
\end{figure}

After performing the Holstein–Primakoff transformation and retaining terms up to quadratic order in the magnon operators, one obtains the quadratic Hamiltonian $\hat{H}_2=\sum_{\bm{k}}\hat{\psi}^{\dagger}_{\bm{k}}H_{\bm{k}}\hat{\psi}_{\bm{k}}$ where $\psi_{\bm{k}}=(\hat{a}_{1,\bm{k}},\hat{a}_{2,\bm{k}},\hat{a}_{3,\bm{k}},\hat{a}_{4,\bm{k}})^{\mathrm{T}}$ is a column vector composed of the magnon annihilation operators for the four sublattices, and $\hat{a}_{\ell,\bm{k}}$ denotes the magnon annihilation operator associated with sublattice $\ell$. The matrix elements of $H_{\bm{k}}$ read
\begin{align}\label{eq:pyro_Ham_elements}
    \left(H_{\bm{k}}\right)_{ij}=
    \begin{cases}
        6JS-g\mu_{\mathrm{B}}B & i=j
        \\[2mm]
        -2S\cos{(\bm{k}\cdot\bm{\delta}_{ij})}\left(J-i\frac{\mathbf{B}\cdot\bm{D}_{ij}}{|\mathbf{B}|}\right) & i\neq j
    \end{cases}.
\end{align}
Here, the divergence of $(H_{\mathbf{k}})_{ij}$ for $\vert \mathbf{B} \vert \rightarrow 0_{-}$ is an artifact of the present model~\cite{Neumann20PRL}. For zero external field, we assume that a weak magnetocrystalline anisotropy pins the moment along $z$, and hence replace $\mathbf{B}$ by $z$ while omitting the diagonal Zeeman shift. The calculated AC-EQ components are shown in Figs.~\ref{fig3}(b)–\ref{fig3}(d).
Since $\mathrm{Lu_2V_2O_7}$ belongs to the magnetic point group $4/mm'm'$, symmetry requires all AC-EQ components to vanish except for $\mathcal{Q}^{(0)}$ and $\mathcal{Q}^{(2)}_{3z^{2}-r^{2}}$. In the absence of the DMI, all AC-EQ components vanish as shown in Fig.~\ref{fig3}(b). This is expected because the DMI introduces complex magnon hopping amplitudes as can be seen from Eq.~(\ref{eq:pyro_Ham_elements}), thereby giving rise to a nontrivial Berry curvature and finite magnon orbital angular momentum, both of which are responsible for the finite AC-EQ in this model [Figs.~\ref{fig3}(b)-\ref{fig3}(d)]. The sign change in Fig.~\ref{fig3}(d) reflects a crossover from the self-rotation-dominated regime at low temperatures to a Berry-curvature-dominated regime at higher temperatures (see Supplemental Material~\cite{suppl_ref}).

Second, we investigate the noncollinear antiferromagnet $\mathrm{Fe}$ langasite, $\mathrm{Ba_3NbFe_3Si_2O_{14}}$, to demonstrate that a finite AC-EQ can arise even in the absence of the DMI. This material exhibits a helical magnetic structure with a magnetic period of approximately seven layers, which we approximate using a commensurate seven-layer magnetic unit cell. The helical magnetic order of $\mathrm{Fe}$ langasite can give rise to a nontrivial magnon Berry curvature and is therefore expected to support a nontrivial magnon OAM texture. Moreover, although the magnon spin magnetic moment texture is absent in this material~\cite{Cheon18PRb}, our calculations reveal a finite magnon orbital magnetic moment texture (see Supplemental Material~\cite{suppl_ref}). Thus, $\mathrm{Fe}$ langasite provides a promising setting in which a finite AC-EQ may emerge even in the absence of the DMI. The effective spin Hamiltonian is given by
\begin{align}
    \hat{H} = \hat{H}_{\mathrm{intra}}+\hat{H}_{\mathrm{inter}}+\hat{H}_{\mathrm{Z}},
\end{align}
where $\hat{H}_{\mathrm{intra}}$, $\hat{H}_{\mathrm{inter}}$ describe the symmetric intralayer and interlayer Heisenberg exchange interaction, respectively, in the absence of the DMI, while $\hat{H}_{\mathrm{Z}}$ denotes the Zeeman interaction. The intralayer and interlayer exchange terms are explicitly given by
\begin{align}
    \hat{H}_{\mathrm{intra}}=\sum_{l}\sum_{\xi\neq\xi',\zeta}\left(J_1\hat{\mathbf{S}}_{l,\xi}\cdot\hat{\mathbf{S}}_{l,\xi'}+J_2\hat{\mathbf{S}}_{l,\xi}\cdot\hat{\mathbf{S}}_{l,\zeta}\right),
\end{align}
\begin{align}
    \hat{H}_{\mathrm{inter}}=\sum_{l}\sum_{\xi,\zeta\neq\zeta'}&\left(J_3\hat{\mathbf{S}}_{l,\xi}\cdot\hat{\mathbf{S}}_{l+1,\zeta'}+J_4\hat{\mathbf{S}}_{l,\xi}\cdot\hat{\mathbf{S}}_{l+1,\xi}\right.\notag\\
    &\left.+J_5\hat{\mathbf{S}}_{l,\xi}\cdot\hat{\mathbf{S}}_{l+1,\zeta}\right),
\end{align}
and the Zeeman terms are given by
\begin{align}
    \hat{H}_{\mathrm{Z}}=-g\mu_{\mathrm{B}}\mathbf{B}\cdot\sum_{l,\xi}\hat{\mathbf{S}}_{l,\xi}.
\end{align}
Here, $l$ labels the magnetic layer, while $\xi$, $\xi'$, $\zeta$, and $\zeta'$ label the magnetic sublattice within each layer. The intralayer and interlayer exchange paths are indicated in Fig.~\ref{fig4}(a). The role of the interlayer Hamiltonian is to select the helical modulation along the $c$ axis. The intralayer exchanges stabilize the $2\pi/3$ order within each Fe triangle, whereas the competing interlayer exchanges $J_3$, $J_4$, and $J_5$ determine the relative azimuthal angle $\tau$ between adjacent triangular layers. Minimizing the classical interlayer energy yields~\cite{Cheon18PRb}
\begin{align}
    \sum_{\xi=0}^{2}J_{3+\xi}\sin{(\tau-\phi_{l,\xi})}=0,
\end{align}
showing that the helical pitch is selected by the symmetric interlayer exchange frustration. Therefore, the interlayer Hamiltonian should be kept explicitly, since it provides the microscopic origin of the $c$ axis helicity and produces the $k_z$-dependent chiral hopping structure in the magnon Hamiltonian.
After performing the Holstein–Primakoff transformation and diagonalizing the resulting magnon Hamiltonian, we calculate the Berry curvature, OAM, and orbital magnetic moment required to evaluate the AC-EQ components. The magnetic structure and the resulting temperature dependence of the AC-EQ components are presented in Fig.~\ref{fig4}. The AC-EM and AC-REQ of Fe langasite also exhibit temperature-dependent sign reversals [Figs.~\ref{fig4}(b) and \ref{fig4}(c)]. These sign changes originate from a crossover between the self-rotation and Berry-curvature contributions, with the former dominating at low temperatures and the latter at higher temperatures (see Supplemental Material~\cite{suppl_ref}). This crossover is explicitly illustrated for the dominant $\mathcal{Q}^{(2)}_{3z^{2}-r^{2}}$ component in Fig.~\ref{fig4}(d).

Even when the crystal axes of a ferromagnetic pyrochlore sample are well aligned, the magnetic moments may form multiple domains with different orientations. In this case, the net magnetic moment can be averaged out over the sample. Since the AC-EQ component $\mathcal{Q}^{(2)}_{3z^2-r^2}$ is tied to the direction of the ferromagnetic order, it can also be averaged out and become difficult to detect. In contrast, in the noncollinear antiferromagnet $\mathrm{Fe}$ langasite, $\mathcal{Q}^{(2)}_{3z^2-r^2}$ is not simply locked to a uniform magnetic moment direction. Instead, it is governed by the noncollinear spin configuration and its relation to the crystallographic chirality. Therefore, once the crystal domains are well aligned, the corresponding $\mathcal{Q}^{(2)}_{3z^2-r^2}$ is not necessarily canceled by the absence of a magnetic domain order. This distinction is experimentally relevant. In ferromagnetic pyrochlores, observing the AC-EQ response may require preparing a single magnetic domain, for example by field cooling or applying a magnetic field. By contrast, in $\mathrm{Fe}$ langasite, the relevant AC-EQ component can remain finite as long as the structural chirality and magnetic helicity are coherently selected. Thus, noncollinear antiferromagnets provide a robust platform for probing AC-EQ responses even when the macroscopic magnetization is compensated.

To put the magnitude into perspective, we convert the electric quadrupole density into an electric quadrupole moment by multiplying it by the unit-cell volume,
\begin{align}
    \mathcal{Q}^{\mathrm{uc}}_{\alpha \beta} = V\mathcal{Q}_{\alpha \beta},
\end{align}
where $V$ is the volume of the unit cell. The resulting values are of order $10^{-9}$--$10^{-8} e\AA^{2}$, or equivalently $10^{-1}$--$1\,e\,\mathrm{b}$, which is comparable to typical nuclear electric quadrupole moments tabulated in barns~\cite{Stone16At}. This comparison is meant only as a scale estimate, since $\mathcal{Q}^{\mathrm{uc}}_{\alpha \beta}$ is an emergent thermodynamic response of magnons rather than a static quadrupole moment of a nuclear charge distribution.

Polarization-resolved Kerr measurements of electronic quadrupoles~\cite{Lee2026arXiv} suggest exploring optical detection. For the dominant $\mathcal{Q}^{(2)}_{3z^2-r^2}$ components [Figs.~\ref{fig3}(d) and~\ref{fig4}(c)], X-ray magnetic linear dichroism provides another possible detection scheme based on electronic quadrupole sum rules~\cite{Yamasaki2025STAM}. Quantitative predictions for these detection schemes require a microscopic calculation connecting the magnon quadrupole to the corresponding optical responses.

The small magnitude of the calculated moments reflects, in part, the bare relativistic AC coupling adopted here, $\widetilde{\mathbf A}_{\mathrm{eff}}=\boldsymbol\mu\times\mathbf E/c^2$, while the magnon Bloch bands and their geometry are included in the calculation. In a solid, however, the effective coupling need not be limited to this bare relativistic scale. For magnons, the possibility of enhanced effective AC gauge coupling is supported theoretically by microscopic studies of electric-field-induced exchange interactions~\cite{Liu2011PRL,Boliasova2025PRb} and experimentally by measurements of electric-field-induced spin-wave phase shifts in YIG~\cite{Zhang2014PRL}. A useful analogy is the electronic Rashba effect, where small interband energy denominators, such as the band gap $E_{\mathrm g}\ll m_e c^2$, can contribute to a substantial enhancement of the effective spin--orbit coupling~\cite{Bahramy11PRb}. A similar mechanism could also enhance the AC-EQ response to electric-field gradients, potentially yielding larger quadrupole responses. Quantifying such material-dependent enhancements is left for future work.

\section{Concluding remarks}\label{sec:conclusion}

In this paper, we have developed a theory of the Aharonov-Casher-induced electric quadrupole (AC-EQ) of charge-neutral particles carrying magnetic moments. The AC-EQ arises from the Aharonov-Casher (AC) coupling, which allows the motion of a magnetic moment to couple to the spatial variation of an electric field without electric charge.

Starting from an intuitive point particle and free-space picture, we demonstrated how  the angular motion of a charge-neutral particle carrying a magnetic moment gives rise to an electric quadrupole. We have then derived the formula of the AC-EQ tensor in crystalline solids using the semiclassical wave-packet formalism. Since the magnetic moment is a vector quantity, the AC-EQ tensor is described by a reducible second-rank Cartesian tensor and contains richer information than the orbital magnetization of electrons. The AC-EQ tensor can be decomposed into the three irreducible components: the AC electric monopole (AC-EM), AC electric toroidal dipole (AC-ETD), and AC reduced electric quadrupole (AC-REQ). These components couple to the divergence, circulation, and shear parts of the electric-field-gradient tensor, respectively. This classification provides a systematic way to identify the symmetrically allowed components of the AC-EQ tensor.

To demonstrate the applicability of our theory, we have investigated two representative magnonic systems: the ferromagnetic pyrochlore, $\mathrm{Lu_2V_2O_7}$ and the noncollinear antiferromagnetic $\mathrm{Fe}$ langasite $\mathrm{Ba_3NbFe_3Si_2O_{14}}$. In $\mathrm{Lu_2V_2O_7}$, Dzyaloshinskii--Moriya interaction (DMI) plays a fundamental role as the relativistic interaction responsible for generating finite AC-EQ components. In the absence of the DMI, all AC-EQ components vanish in this material. On the other hand, the finite AC-EQ responses can also arise from helical magnetic order in $\mathrm{Ba_3NbFe_3Si_2O_{14}}$. These results demonstrate that DMI is not a necessary condition for AC-EQ components, but rather one possible microscopic route to them.

The present formalism focuses on the equilibrium thermodynamic response induced by the AC effect. A natural future direction is to extend it to nonequilibrium regimes, where time-dependent driving force may give rise to dynamical AC-induced electric multipoles. Another complementary direction is to extend the theory to the Bose-Einstein-condensed regime, which is not considered here. Charge-neutral Bosonic quasiparticles carrying magnetic moments, such as magnons and excitons, can form Bose-Einstein-condensed phases under suitable conditions. Recent studies have investigated how the AC effect influences the collective motion and circulation of charge-neutral Bosons in this regime~\cite{Yamamoto25CP,Yamamoto26PRb}. It would be interesting to examine how condensate coherence and supercurrents modify the AC-EQ and whether they generate contributions beyond those captured by the present formulation.

Our findings open a new perspective on the orbital motion of charge-neutral particles and quasiparticles carrying magnetic moments. Although we have demonstrated the theory using magnons as a representative platform, the same principle may apply to a broader class of charge-neutral particles or quasiparticles carrying magnetic dipole moments. Therefore, the AC-EQ offers a route to characterize the orbital motion of charge-neutral magnetic excitations through electric-field-gradient responses rather than ordinary magnetic-field responses.

\begin{acknowledgments}
We thank Jongjun M. Lee, Insu Baek, Seungyun Han, Jeongwoo Kim, and Rhea Hoyer for fruitful discussions. Especially, we are grateful to Rhea Hoyer for useful comments for numerical calculations. H.-W. Lee, S. Cheon, H. Lee, and Y. Jeon were financially supported by the National Research Foundation of Korea (NRF) grant funded by the Korean government (MSIT) (No. RS-2024-00356270 and No. RS-2024-00410027).
\end{acknowledgments}

\appendix

\section{Derivation of the electric quadrupoles}
\label{appendix:Derivation_semiclassical}

In this appendix, we present details of the derivation of the AC-EQ tensor [Eq.~\eqref{eq:AC-EQ_tensor}] based on the semiclassical formalism. We first obtain the wave-packet energy correction generated by the AC effective magnetic field and then evaluate the thermodynamic response of the grand potential to the electric-field gradient to obtain the AC-EQ tensor. In the semiclassical derivation below, $\bm{\mu}_{n\mathbf{k}}$ is understood as the band-projected magnetic moment evaluated in equilibrium. For simplicity, we omit the momentum and band indices unless they are needed.

\subsection{Semiclassical dynamics of the wave packet of the neutral particle}\label{appendix:Semiclassical_dynamics}

We begin with the Hamiltonian modified by the effective electromagnetic potentials $\widetilde{\phi}_{\textrm{eff}}$ and $\widetilde{\textbf{A}}_{\textrm{eff}}$ as
\begin{align}\label{eq:appendix_Hamiltonian}
    \hat{H} &= \hat{H}_{0}\left( \textbf{q} - \frac{1}{\hbar}\widetilde{\textbf{A}}_{\textrm{eff}}(\textbf{r}) \right) + \widetilde{\phi}_{\textrm{eff}}(\textbf{r}).
\end{align}
Then, the local Hamiltonian has the form
\begin{align}\label{eq:appendix_local_Hamiltonian}
    \hat{H}_{\textrm{c}} &= \hat{H}_{0}\left( \textbf{q} - \frac{1}{\hbar}\widetilde{\textbf{A}}_{\textrm{eff}} (\textbf{r}_{\textrm{c}}) \right) + \widetilde{\phi}_{\textrm{eff}}(\textbf{r}_{\textrm{c}}).
\end{align}
As $\widetilde{\textbf{A}}_{\textrm{eff}}(\textbf{r}_{\textrm{c}})$ is only an additive constant to the canonical wave vector $\textbf{q}$, the basis states have the form $\ket{u (\textbf{r}_{\textrm{c}},\textbf{q})} = \ket{u(\textbf{k})}$, where $\textbf{k} = \textbf{q} - \widetilde{\textbf{A}}_{\textrm{eff}}(\textbf{r}_{\textrm{c}})/\hbar$ is the kinematic wave vector.

The eigenenergy can be written in the form
\begin{align}\label{eq:appendix_local_energy}
    \mathcal{E}_{\textrm{c}}(\textbf{r}_{\textrm{c}},\textbf{k}) &= \mathcal{E}_{0}(\textbf{k}) + \widetilde{\phi}_{\textrm{eff}}(\textbf{r}_{\textrm{c}}),
\end{align}
where $\mathcal{E}_{0}(\textbf{k}) = \mathcal{E}_{n\textbf{k}}$ is the energy eigenvalue of $\hat{H}_{0}$.
The gradient correction to the energy has the form
\begin{align}\label{eq:appendix_energy_correction_raw}
    \Delta \mathcal{E} &= - \textrm{Im} \left[ \left\langle \frac{\partial u}{\partial \textbf{r}_{\textrm{c}}} \biggr\vert \cdot (\mathcal{E}_{\textrm{c}} - \hat{H}_{\textrm{c}}) \biggr\vert \frac{\partial u}{\partial \textbf{q}} \right\rangle \right],
\end{align}
to first order in spatial gradients~\cite{Sundaram99PRb}. Using the chain rule, we replace the derivative of $\textbf{r}_{\textrm{c}}$ and $\textbf{q}$ with the derivative of $\textbf{k}$. Here, we neglect the $\mathbf{q}$-dependence of $\widetilde{\mathbf{A}}_{\text{eff}}$ because it contributes only through products such as $(\partial_{\textbf{r}_{\text{c}}}\widetilde{A}_{\text{eff},\alpha})(\partial_{\mathbf{q}}\widetilde{A}_{\text{eff},\beta})$, which are second order in the external perturbations. Substituting these relations into Eq. \eqref{eq:appendix_energy_correction_raw}, we obtain
\begin{align}\label{eq:appendix_energy_correction}
    \Delta \mathcal{E} &= -\textbf{m}_{\textrm{E}} \cdot \widetilde{\textbf{B}}_{\textrm{eff}},
\end{align}
where
\begin{align}\label{eq:appendix_small_me}
    \textbf{m}_{\textrm{E}} &= \frac{1}{2\hbar} \, \textrm{Im} \left[ \braket{\nabla_{\textbf{k}}u | \times (\mathcal{E}_{0}  - \hat{H}_{0}(\textbf{k})) | \nabla_{\textbf{k}}u} \right]_{\textbf{k} = \textbf{k}_{c}}
\end{align}
is the orbital AC moment of the Bloch states.

Finally, we present the Lagrangian and the equations of motion of the wave packet. The Lagrangian has the form
\begin{align}\label{eq:appendix_Lagrangian}
    \mathcal{L} &= -\mathcal{E}_{m} - \widetilde{\phi}_{\textrm{eff}}(\textbf{r}_{\textrm{c}}) + \hbar \dot{\textbf{r}}_{\textrm{c}} \cdot \textbf{k}_{\textrm{c}}  + \dot{\textbf{r}}_{\textrm{c}} \cdot \widetilde{\textbf{A}}_{\textrm{eff}}(\textbf{r}_{\textrm{c}})\notag
    \\[2mm]
    &\quad + \hbar \dot{\textbf{k}}_{\textrm{c}} \cdot i\left\langle u \biggr\vert \frac{\partial u}{\partial \textbf{k}_{\textrm{c}}} \right\rangle,
\end{align}
where $\mathcal{E}_{m} \equiv \mathcal{E}_{0} (\textbf{k}_{\textrm{c}}) - \textbf{m}_{\textrm{E}} \cdot \widetilde{\textbf{B}}_{\textrm{eff}}$ is the energy corrected by the orbital AC moment. The semiclassical equations of motion are obtained from the Euler-Lagrange equations as
\begin{align}\label{eq:appendix_EOM}
    \dot{\textbf{r}}_{\textrm{c}} &= \frac{1}{\hbar} \nabla_{\textbf{k}_{\textrm{c}}} \mathcal{E}_{m} - \dot{\textbf{k}}_{\textrm{c}} \times \bm{\Omega},
    \\[2mm]
    \hbar \dot{\textbf{k}}_{\textrm{c}} &= \widetilde{\mathbf{E}}_{\text{eff}} + \dot{\textbf{r}_{\textrm{c}}} \times \widetilde{\mathbf{B}}_{\text{eff}}.
\end{align}

\subsection{The Aharonov-Casher-induced electric quadrupole}\label{appendix:AC_quadrupole}

The effective magnetic field $\widetilde{\mathbf{B}}_{\text{eff}}$ modifies the volume element of the phase space $\Delta V = \Delta \textbf{r} \Delta \textbf{k}$, which obeys the equation of motion~\cite{Xiao05PRL}
\begin{align}\label{eq:appendix_EOM_phase_volume}
    \frac{1}{\Delta V}\frac{d \Delta V}{dt} &= \nabla_{\textbf{r}} \cdot \dot{\textbf{r}} + \nabla_{\textbf{k}} \cdot \dot{\textbf{k}}.
\end{align}
From this equation, we can obtain the relation between the unperturbed volume element $\Delta V_{0}$ and the modified volume element due to the external effective fields,
\begin{align}\label{eq:appendix_modified_volume}
    \Delta V &= \frac{\Delta V_{0}}{(1 - \widetilde{\textbf{B}}_{\textrm{eff}}\cdot \bm{\Omega}/\hbar)}.
\end{align}
The modified density of states is defined such that the number of states in the volume element, $\mathcal{D}_{n}(\textbf{r}, \textbf{k}) \Delta V$, remains constant in time as
\begin{align}\label{eq:appendix_modified_DOS}
    \mathcal{D}_{n} &= (2\pi)^{-d}(1 - \frac{1}{\hbar}\widetilde{\textbf{B}}_{\textrm{eff}} \cdot \bm{\Omega}_{n}).
\end{align}
The real-space density of a physical observable $\hat{O}$ is written as
\begin{align}\label{eq:appendix_observable}
    \bar{O} &\approx \sum_{n} \int d\textbf{k} \, \mathcal{D}_{n}(\textbf{k}) f_{n\mathbf{k}} O_{n}(\textbf{k}),
\end{align}
where $O_{n}(\textbf{k})$ is the expectation value of $\hat{O}$ in a Bloch state, and
\begin{align}\label{eq:appendix_occupation_function_approx}
    f_{n\mathbf{k}} &\approx F_{n\mathbf{k}}
    + \beta F_{n\mathbf{k}}(1\pm F_{n\mathbf{k}})\textbf{m}_{\textrm{E},n\mathbf{k}} \cdot \widetilde{\textbf{B}}_{\textrm{eff},n\mathbf{k}}
\end{align}
is the distribution function.
Here, $F \equiv 1/(e^{\beta (\mathcal{E}_{0} - \mu_{\textrm{ch}})} \mp 1)$ is the Bose-Einstein (upper sign) or Fermi-Dirac distribution (lower sign), respectively, $\beta = 1/k_{\text{B}}T$, and $\mu_{\textrm{ch}}$ is the chemical potential. Then, the total energy of the system is written as
\begin{align}\label{eq:appendix_total_energy}
    \mathcal{E}_{\textrm{tot}} &= \sum_{n} \int d\textbf{k}\, f_{n\textbf{k}}\mathcal{D}_{n\textbf{k}}\left(\mathcal{E}_{n\textbf{k}} - \textbf{m}_{\textrm{E},n\textbf{k}} \cdot \widetilde{\textbf{B}}_{\textrm{eff},n\textbf{k}}\right) \notag
    \\[2mm]
    &\approx \sum_{n} \int \frac{d\textbf{k}}{(2\pi)^{d}} \left[ F_{n\textbf{k}} + \frac{\beta}{c^{2}} F_{n\textbf{k}}(1\pm F_{n\textbf{k}}) \mathbf{m}_{\text{E},n\textbf{k}} \cdot \bm{\Gamma}_{n\textbf{k}} \right] \notag
    \\[2mm]
    &\quad \times \left( 1 - \frac{1}{\hbar c^{2}} \bm{\Omega}_{n\textbf{k}} \cdot \bm{\Gamma}_{n\textbf{k}}\right) \left( \mathcal{E}_{n\textbf{k}} - \frac{1}{c^{2}} \mathbf{m}_{E,n\textbf{k}} \cdot \bm{\Gamma}_{n\textbf{k}} \right),
\end{align}
where
\begin{align}\label{eq:Gamma}
    \bm{\Gamma} &= (\mathbf{E} \cdot \nabla ) \bm{\mu} + \bm{\mu}(\nabla \cdot \mathbf{E}) - \mathbf{E}(\nabla \cdot \bm{\mu}) - (\bm{\mu} \cdot \nabla )\mathbf{E}.
\end{align}
We emphasize that $\nabla$ is the spatial derivative, not the $\mathbf{k}$-derivative $\nabla_{\mathbf{k}}$.

Now, we evaluate the linear responses of charge-neutral particles. The linear response of the total energy to the electric field and the spatial derivative of the electric field are calculated by
\begin{widetext}
\begin{align}\label{eq:appendix_E_derivative-E}
    \frac{\partial \mathcal{E}_{\textrm{tot}}}{\partial E_{\alpha}} = \frac{1}{c^{2}}\sum_{n}\int \frac{d\textbf{k}}{(2\pi)^{d}} \sum_{\rho}&\biggr[ F\left\{ \beta \mathcal{E}_{0} (1 \pm F) - 1\right\} m_{E,\rho} - \frac{\mathcal{E}_{0}F}{\hbar}\Omega_{\rho} \biggr] \left( \partial_{\alpha}\mu_{\rho} - \delta_{\alpha\rho}\sum_{\lambda}\partial_{\lambda}\mu_{\lambda}\right) ,
\end{align}
\begin{align}\label{eq:appendix_E_derivative-partialE}
    \frac{\partial \mathcal{E}_{\textrm{tot}}}{\partial (\partial_{\beta} E_{\alpha})} = \frac{1}{c^{2}} \sum_{n} \int \frac{d\textbf{k}}{(2\pi)^{d}} \sum_{\rho} &\left( \mu_{\rho} \delta_{\alpha \beta} - \mu_{\beta}\delta_{\rho\alpha}\right) \left[ F \left( \beta \mathcal{E}_{0}(1 \pm F) - 1 \right) m_{E,\rho} - \frac{\mathcal{E}_{0}F}{\hbar}\Omega_{\rho} \right] .
\end{align}
\end{widetext}
Here, we omit the indices $n$ and $\mathbf{k}$.

Next, we consider the entropy $S$ of the system. The entropy has the form
\begin{align} \label{eq:entropy}
    & S = -k_\textrm{B}\sum_{n}\int d\textbf{k} \, \mathcal{D}_n \left[ f_n\ln{f_n} \mp (1 \pm f_n)\ln{(1 \pm f_n)} \right].
\end{align}
We obtain the linear responses of $TS$ to the electric field and the spatial derivative of the electric field as follows:
\begin{widetext}
\begin{align}\label{eq:appendix_TS_derivative-E}
    \frac{\partial (TS)}{\partial E_\alpha} = \frac{1}{c^{2}}\sum_{n}\int \frac{d\textbf{k}}{(2\pi)^{d}} \, &\sum_{\rho}\biggr[\beta(\mathcal{E}_0 - \mu_{\textrm{ch}})F(1 \pm F)m_{E,\rho} -\frac{1}{\hbar\beta}\left\{ \beta(\mathcal{E}_0-\mu_{\textrm{ch}}) F \mp \ln{\left( 1 \mp e^{-\beta (\mathcal{E}_{0}-\mu_{\textrm{ch}})} \right)} \right\}\Omega_{\rho} \biggr] \notag
    \\[2mm]
    &\times \left( \partial_{\alpha}\mu_{\rho} - \delta_{\alpha\rho}\sum_{\lambda}\partial_{\lambda}\mu_{\lambda}\right)
\end{align}
\begin{align}\label{eq:appendix_TS_derivative-partialE}
    \frac{\partial (TS)}{\partial (\partial_\beta E_\alpha)} = \frac{1}{c^{2}} \sum_{n} \int \frac{d\textbf{k}}{(2\pi)^{d}} \, &\sum_{\rho}\left( \mu_{\rho} \delta_{\alpha \beta} - \mu_{\beta}\delta_{\rho\alpha}\right) \notag
    \\
    &\times \left[ \beta (\mathcal{E}_{0}-\mu_{\textrm{ch}})F(1 \pm F) m_{E,\rho} - \frac{1}{\hbar \beta} \left\{ \beta (\mathcal{E}_{0}-\mu_{\textrm{ch}}) F \mp \ln(1 \mp e^{-\beta (\mathcal{E}_{0}-\mu_{\textrm{ch}})})  \right\} \Omega_{\rho} \right] .
\end{align}
\end{widetext}
The grand potential $G$ is given by
\begin{align}\label{eq:appendix_grand_potential}
    G = \mathcal{E}_{\text{tot}} - \mu_{\text{ch}}\mathcal{N} - TS,
\end{align}
where
\begin{align}\label{eq:appendix_number}
    \mathcal{N} &= \sum_{n}\int d\textbf{k} \, f_n\mathcal{D}_n
\end{align}
is the particle (or quasiparticle) number.
The linear responses of $\mathcal{N}$ to the electric field and the spatial derivative of the electric field are given by
\begin{widetext}
\begin{align}\label{eq:appendix_N_derivative-E}
    \frac{\partial \mathcal{N}}{\partial E_{\alpha}} = \frac{1}{c^{2}} \sum_{n} \int \frac{d\textbf{k}}{(2\pi)^{d}} &\sum_{\rho}\biggr[ \beta F(1 \pm F) m_{E,\rho} - \frac{F}{\hbar}\Omega_{\rho} \biggr] \left( \partial_{\alpha}\mu_{\rho} - \delta_{\alpha\rho}\sum_{\lambda}\partial_{\lambda}\mu_{\lambda}\right) ,
\end{align}
\begin{align}\label{eq:appendix_N_derivative-partialE}
    \frac{\partial \mathcal{N}}{\partial (\partial_{\beta} E_{\alpha})} = \frac{1}{c^{2}} \sum_{n} \int \frac{d\textbf{k}}{(2\pi)^{d}} &\sum_{\rho}\left( \mu_{\rho} \delta_{\alpha \beta} - \mu_{\beta}\delta_{\rho\alpha}\right)  \left[ F \beta (1 \pm F) m_{E,\rho} - \frac{F}{\hbar}\Omega_{\rho} \right] .
\end{align}
\end{widetext}
Therefore, we obtain both the linear responses of the grand potential to the electric field and the spatial derivative of the electric field as follows:
\begin{widetext}
\begin{align}\label{eq:appendix_AC-polarization}
    \mathcal{P}_{\alpha} \equiv -\frac{\partial G}{\partial E_\alpha} = \frac{1}{c^2} \sum_{n} \int \frac{d\textbf{k}}{(2\pi)^{d}} \sum_{\rho}\, \biggr[& Fm_{E,\rho} \pm \frac{1}{\hbar\beta}\Omega_\rho\ln{(1 \mp e^{-\beta(\mathcal{E}_{0}-\mu_{\textrm{ch}})})} \biggr] \left( \partial_{\alpha}\mu_{\rho} - \delta_{\alpha\rho}\sum_{\lambda}\partial_{\lambda}\mu_{\lambda}\right),
\end{align}
\begin{align}\label{eq:appendix_AC-electric_quadrupole}
    \mathcal{Q}_{\alpha \beta} \equiv -\frac{\partial G}{\partial (\partial_\beta E_\alpha)} &= \frac{1}{c^{2}} \sum_{n} \int \frac{d\textbf{k}}{(2\pi)^{d}} \sum_{\rho} \left( \mu_{\rho} \delta_{\alpha \beta} - \mu_{\beta}\delta_{\rho\alpha}\right) \left[ Fm_{E,\rho} \pm \frac{1}{\hbar \beta}\Omega_{\rho}\ln{\left( 1 \mp e^{-\beta (\mathcal{E}_{0}-\mu_{\textrm{ch}})} \right)} \right] .
\end{align}
\end{widetext}
First, we focus on Eq.~\eqref{eq:appendix_AC-electric_quadrupole}, which is equivalent to Eq.~\eqref{eq:AC-EQ_tensor}. Introducing the $\bm{\Lambda}$ [Eq.~\eqref{eq:Lambda}], Eq.~\eqref{eq:appendix_AC-electric_quadrupole} can be rewritten as Eq.~\eqref{eq:AC-EQ_tensor}. $\mathcal{Q}_{\alpha \beta}$ can be interpreted as the electric quadrupole induced by the angular motion of charge-neutral particles through the AC effect. This formula is a relativistic counterpart to the orbital magnetization of electrons
\begin{align}\label{eq:appendix_OM_electron}
    M_{\alpha} &= e\sum_{n}\int \frac{d\textbf{k}}{(2\pi)^{d}} \left[ \frac{F^{\textrm{FD}}}{e}m_{\alpha} + \frac{1}{\hbar \beta}\Omega_{\alpha} \textrm{ln} \left( 1 + e^{-\beta (\mathcal{E}_{0}-\mu_{\textrm{ch}})} \right) \right],
\end{align}
where $F^{\text{FD}} = 1/(e^{\beta (\mathcal{E}_{0} - \mu_{\textrm{ch}})} + 1)$ is the Fermi-Dirac distribution~\cite{Shi07PRL}.

Next, we focus on Eq.~\eqref{eq:appendix_AC-polarization}. This can be interpreted as the polarization induced by the angular motion of charge-neutral particles through the AC effect. Unlike the AC-EQ, the polarization $\mathcal{P}_{\alpha}$ is finite only in the presence of a spatial texture of $\bm{\mu}$. In this work, to focus on $\mathcal{Q}_{\alpha \beta}$, we ignore the spatial texture of $\bm{\mu}$ and consider only the momentum-space texture, leaving $\mathcal{P}_{\alpha}$ for future work.

This derivation assumes weak and slowly varying external fields, keeps only terms linear in the electric-field gradient, and uses the single-band adiabatic wave-packet approximation. In the presence of band degeneracies, the corresponding non-Abelian generalization is required~\cite{Cheng13PRb}.

\section{Calculation details}\label{Appendix:calculation_details}

\subsection{Berry curvature and orbital moment in Bosonic systems}

Consider a generic Bosonic Hamiltonian
\begin{align}
    \hat{\mathcal{H}} &= \frac{1}{2}\sum_{\mathbf{k}} \bm{\Psi}_{\mathbf{k}}^{\dagger} H_{\mathbf{k}} \bm{\Psi}_{\mathbf{k}},
\end{align}
where the Nambu spinors are $\bm{\Psi}_{\mathbf{k}} = [\bm{\beta}_{\mathbf{k}} \; \bm{\beta}_{-\mathbf{k}}^{\dagger}]^{\textrm{T}}$ and $\bm{\Psi}_{\mathbf{k}}^{\dagger} = [\bm{\beta}^{\dagger}_{\mathbf{k}} \; \bm{\beta}_{-\mathbf{k}}]$, the Bosonic creation operators are $\bm{\beta}_{\mathbf{k}}^{\dagger} = [\beta_{1,\mathbf{k}}^{\dagger}, ..., \beta_{N,\mathbf{k}}^{\dagger}]$. Here, $N$ is the number of internal degrees of freedom considered within a unit cell. The Nambu spinor satisfies the Bosonic commutation relation $[\bm{\Psi}_{\mathbf{k}},\bm{\Psi}_{\mathbf{k}^{\prime}}^{\dagger}] = \sigma_{3} \delta_{\mathbf{k}\mathbf{k}^{\prime}}$, where $\sigma_{3} = \textrm{diag}[I_{(N)}, -I_{(N)}]$ (i.e., $[\sigma_{3}]_{nm} = s_{n}\delta_{nm}$ with $s_{n} = +1$ for $n \in \{1, ..., N\}$ and $s_{n} = -1$ for $n \in \{N+1, ..., 2N\}$). The Bosonic Bogoliubov-de Gennes (BdG) Hamiltonian $H_{\mathbf{k}}$ is a $2N \times 2N$ Hermitian matrix of the form
\begin{align}\label{eq:appendix_BdG_matrix}
    H_{\mathbf{k}} &= \left[
    \begin{array}{cc}
        \mathbf{a}_{\mathbf{k}} & \mathbf{b}_{\mathbf{k}} \\
        \mathbf{b}_{-\mathbf{k}}^{\ast} & \mathbf{a}_{-\mathbf{k}}^{\ast}
    \end{array}\right],
\end{align}
where $\mathbf{a}_{\mathbf{k}}$ and $\mathbf{a}_{-\mathbf{k}}^{\ast}$ are $N \times N$ normal parts (particle-hole channel) and $\mathbf{b}_{\mathbf{k}}$ and $\mathbf{b}_{-\mathbf{k}}^{\ast}$ are $N \times N$ anomalous parts (particle-particle channel). Now, we introduce a new Nambu basis $\bm{\Phi}_{\mathbf{k}} = V_{\mathbf{k}}^{-1}\bm{\Psi}_{\mathbf{k}} = [\gamma_{\mathbf{k}} \; \gamma_{-\mathbf{k}}^{\dagger}]^{\textrm{T}}$ for the Bogoliubov quasiparticles, such that
\begin{align}
    \bm{\Psi}_{\mathbf{k}}^{\dagger} H_{\mathbf{k}} \bm{\Psi}_{\mathbf{k}} = \bm{\Phi}_{\mathbf{k}}^{\dagger} \mathcal{E}_{\mathbf{k}} \bm{\Phi}_{\mathbf{k}},
\end{align}
where $\mathcal{E}_{\mathbf{k}} = \textrm{diag}[\mathcal{E}_{1,\mathbf{k}}, ..., \mathcal{E}_{N,\mathbf{k}}, \mathcal{E}_{1,-\mathbf{k}}, ... \mathcal{E}_{N,-\mathbf{k}}]$.
Note that $V_{\mathbf{k}}\bm{\Phi}_{\mathbf{k}} = \bm{\Psi}_{\mathbf{k}}$ and $\bm{\Phi}_{\mathbf{k}}^{\dagger}V^{\dagger}_{\mathbf{k}} = \bm{\Psi}^{\dagger}_{\mathbf{k}}$. That is, $V_{\mathbf{k}}$ diagonalizes the BdG Hamiltonian as follows:
\begin{align}
    V_{\mathbf{k}}^{\dagger} H_{\mathbf{k}} V_{\mathbf{k}} &= \mathcal{E}_{\mathbf{k}} = \left[
    \begin{array}{cc}
        \mathcal{E}_{\mathbf{k}}^{+} &  \\
         & \mathcal{E}_{\mathbf{k}}^{-}.
    \end{array} \right]
\end{align}
Here, $\mathcal{E}_{\mathbf{k}}^{+} = \textrm{diag}[\mathcal{E}_{1,\mathbf{k}}, ..., \mathcal{E}_{N,\mathbf{k}}]$ and $\mathcal{E}_{\mathbf{k}}^{-} = \textrm{diag}[\mathcal{E}_{1,-\mathbf{k}}, ... \mathcal{E}_{N,-\mathbf{k}}] = \mathcal{E}^{+}_{-\mathbf{k}}$. Since the new Nambu spinors must satisfy the Bosonic commutation relation (i.e., $[\bm{\Psi}_{\mathbf{k}},\bm{\Psi}_{\mathbf{k}^{\prime}}^{\dagger}] = \sigma_{3} \delta_{\mathbf{k}\mathbf{k}^{\prime}}$), $V_{\mathbf{k}}$ is paraunitary and satisfies
\begin{align}
    V_{\mathbf{k}}^{\dagger} \sigma_{3} V_{\mathbf{k}} &= V_{\mathbf{k}} \sigma_{3} V_{\mathbf{k}}^{\dagger} = \sigma_{3}.
\end{align}
Obviously, the inverse of $V_{\mathbf{k}}$ is given by $V_{\mathbf{k}}^{-1} = \sigma_{3}V_{\mathbf{k}}^{\dagger}\sigma_{3}$. Using this relation, we obtain
\begin{align}\label{eq:appendix_Bogoliubov}
    V_{\mathbf{k}}^{-1}\sigma_{3}H_{\mathbf{k}}V_{\mathbf{k}} = \bar{\mathcal{E}}_{\mathbf{k}},
\end{align}
where
\begin{align}\label{eq:appendix_bar_E}
    \bar{\mathcal{E}}_{\mathbf{k}} = \sigma_{3}\mathcal{E}_{\mathbf{k}} = \left[
    \begin{array}{cc}
        \mathcal{E}_{\mathbf{k}}^{+} &  \\
         & -\mathcal{E}_{\mathbf{k}}^{-}
    \end{array}\right].
\end{align}

The paraunitary matrix $V_{\mathbf{k}}$ can be obtained by Colpa's method~\cite{Colpa78Physa}. We first perform a Cholesky decomposition to find an upper triangular matrix $K_{\mathbf{k}}$ such that $H_{\mathbf{k}} = K^{\dagger}_{\mathbf{k}}K_{\mathbf{k}}$. If $H_{\mathbf{k}}$ is positive definite, then this decomposition is allowed and also $K_{\mathbf{k}}^{-1}$ exists. Next, we introduce a unitary matrix $U_{\mathbf{k}}$ which diagonalizes a Hermitian matrix $W_{\mathbf{k}} = K_{\mathbf{k}}\sigma_{3}K_{\mathbf{k}}^{\dagger}$:
\begin{align}
    U^{\dagger}_{\mathbf{k}}W_{\mathbf{k}}U_{\mathbf{k}} = \bar{\mathcal{E}}_{\mathbf{k}}.
\end{align}
If $H_{\mathbf{k}}$ is positive definite, then the Sylvester's law of inertia guarantees that both $\mathcal{E}_{\mathbf{k}}^{+}$ and $\mathcal{E}_{\mathbf{k}}^{-}$ can be constructed as positive definite diagonal matrices. We can construct $V_{\mathbf{k}}$ as
\begin{align}
    V_{\mathbf{k}} = K^{-1}_{\mathbf{k}}U_{\mathbf{k}}\left[
    \begin{array}{cc}
        (\mathcal{E}_{\mathbf{k}}^{+})^{1/2} &  \\
         & (\mathcal{E}_{\mathbf{k}}^{-})^{1/2}
    \end{array}\right].
\end{align}

Now, we introduce the right energy eigenstates $\{ \ket{u^{\textrm{R}}_{n\mathbf{k}}} \}$, which are represented by $2N \times 1$ column vectors, such that $V_{\mathbf{k}} = \left( \ket{u^{\textrm{R}}_{1\mathbf{k}}} \ket{u^{\textrm{R}}_{2\mathbf{k}}} \cdots \ket{u^{\textrm{R}}_{2N\mathbf{k}}} \right)$. We also introduce the left energy eigenstates $\{ \bra{u^{\textrm{L}}_{n\mathbf{k}}} \}$, which are represented by $1 \times 2N$ row vectors, such that $V_{\mathbf{k}}^{\dagger}\sigma_{3}= \left( \ket{u^{\textrm{L}}_{1\mathbf{k}}} \ket{u^{\textrm{L}}_{2\mathbf{k}}} \cdots \ket{u^{\textrm{L}}_{2N\mathbf{k}}} \right)^{\textrm{T}}$. Let us denote $\ket{u_{n\mathbf{k}}} = \ket{u_{n\mathbf{k}}^{\textrm{R}}}$. Then, we have the relations
\begin{align}
    \bra{u_{n}^{\textrm{L}}} = \bra{u_{n}^{\textrm{R}}}\sigma_{3} &= \bra{u_{n\mathbf{k}}}\sigma_{3}\notag
    \\[2mm]
    \braket{u_{n\mathbf{k}}^{\textrm{L}}|u_{m\mathbf{k}}^{\textrm{R}}} &= \braket{u_{n\mathbf{k}}|\sigma_{3}|u_{m\mathbf{k}}} = \sigma_{nm} = s_{n}\delta_{nm}.
\end{align}
The completeness relation is given by $\sum_{n}s_{n}\ket{u_{n\mathbf{k}}^{\textrm{R}}}\bra{u_{n\mathbf{k}}^{\textrm{L}}} = \mathbb{1}$. From Eq.~\eqref{eq:appendix_Bogoliubov}, we also have $\sigma_{3}H_{\mathbf{k}}\ket{u_{n\mathbf{k}}} = \bar{\mathcal{E}}_{n\mathbf{k}}\ket{u_{n\mathbf{k}}}$, where $\bar{\mathcal{E}}_{n\mathbf{k}}$ is the $n$th diagonal element of $\bar{\mathcal{E}}_{\mathbf{k}}$. The projection operator onto the $n$th band is given by
\begin{align}
    P_{n} = V_{\mathbf{k}}\Gamma_{n}V^{-1}_{\mathbf{k}} = V_{\mathbf{k}}\Gamma_{n}\sigma_{3}V_{\mathbf{k}}^{\dagger}\sigma_{3} = s_{n}\ket{u_{n\mathbf{k}}}\bra{u_{n\mathbf{k}}}\sigma_{3},
\end{align}
where $\Gamma_{n}$ is a diagonal matrix taking $+1$ for the $n$th diagonal component and zero otherwise~\cite{Shindou13PRb}. This projection operator has the properties: $\sum_{n} P_{n}=\mathbb{1}$ and $P_{n} P_{m} = \delta_{nm}P_{n}$.

We introduce the three gauge-covariant objects which consist of the projection operators:
\begin{subequations}
\begin{align}
    \mathcal{F}_{n,\mu\nu} &= \textrm{tr}[(\partial_{\mu}P_{n})Q_{n}(\partial_{\nu}P_{n})],
\end{align}
\begin{align}
    \mathcal{G}_{n,\mu\nu} &= \textrm{tr}[(\partial_{\mu}P_{n})Q_{n}\sigma_{3}H_{\mathbf{k}}Q_{n}(\partial_{\nu}P_{n})],
\end{align}
\begin{align}
    \mathcal{K}_{n,\mu\nu} &= \mathcal{E}_{n}\textrm{tr}[(\partial_{\mu}P_{n})Q_{n}(\partial_{\nu}P_{n})].
\end{align}
\end{subequations}
Here, $\textrm{tr}[A] = \sum_{n=1}^{2N}s_{n}\braket{u_{n}^{\textrm{L}}|A|u_{n}^{\textrm{R}}}$ and $Q_{n} = \mathbb{1} - P_{n}$. Equivalently, these objects can be expressed in terms of $\ket{u_{j}}$ as follows:
\begin{subequations}
\begin{align}
    \mathcal{F}_{n,\mu\nu} &= s_n\braket{\partial_{\mu}u_{n}|\sigma_{3}Q_{n}|\partial_{\nu}u_{n}},
\end{align}
\begin{align}
    \mathcal{G}_{n,\mu\nu} &= s_n\braket{\partial_{\mu}u_{n}|\sigma_{3}Q_{n}\sigma_{3}H_{\mathbf{k}}Q_{n}|\partial_{\nu}u_{n}},
\end{align}
\begin{align}
    \mathcal{K}_{n,\mu\nu} &= s_n\mathcal{E}_{n}\braket{\partial_{\mu}u_{n}|\sigma_{3}Q_{n}|\partial_{\nu}u_{n}}.
\end{align}
\end{subequations}

Then, $\mathbf{m}_{\text{E},n}$ and $\bm{\Omega}_{n}$ are expressed as
\begin{align}
    m_{\text{E},n,\lambda} &= -\frac{1}{2\hbar} \text{Im} \, \left[ \varepsilon_{\lambda \mu \nu} (\mathcal{G}_{n, \mu \nu} - \mathcal{K}_{n, \mu \nu}) \right],
\end{align}
and
\begin{align}
    \Omega_{n,\lambda} &= -\text{Im} \, \left[ \varepsilon_{\lambda \mu \nu} \mathcal{F}_{n, \mu \nu} \right],
\end{align}
respectively. We note that physical $\mathbf{m}_{\text{E},n}$ and $\bm\Omega_{n}$ are evaluated for the physical bands $n = 1, ..., N$.

\section{Zero-point contribution to the AC-EQ}\label{Appendix:ZP}

In this section, we discuss whether quantum zero-point fluctuations generate an additional temperature-independent contribution to the AC-EQ for magnonic systems. In linear spin-wave theory~\cite{Neumann20PRL}, a Bosonic 
BdG Hamiltonian can be written in the Nambu representation as
\begin{align}
    \hat{H} &= \mathcal{E}_{\text{cl}} + \mathcal{E}_{\text{ZP}} + \sum_{\mathbf{k}}\sum_{n=1}^{N}\mathcal{E}_{n,\mathbf{k}} \hat{\gamma}_{n\mathbf{k}}^{\dagger}\hat{\gamma}_{n\mathbf{k}},
\end{align}
where $\mathcal{E}_{\text{cl}}$ is the classical ground state energy, $\mathcal{E}_{\text{ZP}}$ is the harmonic zero-point quantum fluctuation energy, $N$ is the number of magnon modes, and $\hat{\gamma}_{n\mathbf{k}}^{(\dagger)}$ are the Bosonic annihilation (creation) operators, respectively. Here, we introduced the harmonic approximation, which neglects Boson-Boson interactions and effectively includes up to quadratic terms. The harmonic zero-point fluctuation energy contains two contributions, $\mathcal{E}_{\text{ZP}} = \mathcal{E}_{\text{ZP}}^{(1)} + \mathcal{E}_{\text{ZP}}^{(2)}$, where
\begin{align}
    \mathcal{E}_{\text{ZP}}^{(1)} &= - \frac{1}{4} \sum_{\mathbf{k}} \text{tr} \left[ H_{\mathbf{k}} \right] , \qquad \mathcal{E}_{\text{ZP}}^{(2)} = \frac{1}{2} \sum_{\mathbf{k}} \sum_{n=1}^{N} \mathcal{E}_{n\mathbf{k}}.
\end{align}
Consider the generic BdG Hamiltonian in Eq.~\eqref{eq:appendix_BdG_matrix}. If the quadratic Hamiltonian conserves the number of Bosons, $\mathbf{b}_{\mathbf{k}} = 0$. Then, the Bogoliubov transformation reduces to an ordinary unitary transformation. In this case, we have
\begin{align}
    \sum_{\mathbf{k},n}\mathcal{E}_{n\mathbf{k}} &= \sum_{\mathbf{k}} \text{tr} \left[ \mathbf{a}_{\mathbf{k}} \right],
\end{align}
which implies that $\mathcal{E}_{\text{ZP}} = \mathcal{E}_{\text{ZP}}^{(1)} + \mathcal{E}_{\text{ZP}}^{(2)} = 0$. Thus, a number-conserving quadratic Bosonic Hamiltonian does not possess a nontrivial harmonic zero-point correction.

On the other hand, if the anomalous terms $\mathbf{b}_{\mathbf{k}}$ are nonzero, then the creation and annihilation operators are mixed by the Bogoliubov transformation and nonvanishing quantum fluctuations are allowed. Therefore, in this case, an explicit zero-point analysis is required. We note that the terminology ``number conserving" introduced in this section refers to the absence of anomalous terms (i.e., $\mathbf{b}_{\mathbf{k}}=0$).

Next, we formulate the response of the zero-point energy to an electric-field gradient. From Eqs.~\eqref{eq:effective_potentials_tilde} and~\eqref{eq:appendix_local_Hamiltonian}, we obtain
\begin{align}\label{eq:appendix_delta_H_SC}
    \delta H &= -\frac{1}{\hbar}\frac{\partial H_{0}}{\partial k_{\alpha}}\widetilde{A}_{\text{eff},\alpha} = -\mathbf{P} \cdot \mathbf{E},
\end{align}
where $\mathbf{P} = \mathbf{v} \times \bm{\mu}/c^{2}$ and $v_{\alpha} = \partial_{k_{\alpha}}H_{0}/\hbar$. To extend Eq.~\eqref{eq:appendix_delta_H_SC} to the momentum-dependent thermodynamic magnetic moment used in the present theory, we define
\begin{align}
    \hat{\mathcal{M}}_{\alpha} (\mathbf{k}) &= -\frac{\partial \hat{H}_{\mathbf{k}}}{\partial B_{\alpha}}
\end{align}
whose diagonal projection gives $\mu_{n\mathbf{k},\alpha} = \braket{u_{n\mathbf{k}}|\hat{\mathcal{M}}_{\alpha}|u_{n\mathbf{k}}}$. A natural Hermitian operator extension of the polarization in Eq.~\eqref{eq:appendix_delta_H_SC} is
\begin{align}
    \hat{\mathcal{P}}_{\alpha} &= \frac{1}{2c^{2}} \varepsilon_{\alpha \beta \gamma} \{ \hat{v}_{\beta}, \hat{\mathcal{M}}_{\gamma}\} .
\end{align}
Then, Eq.~\eqref{eq:appendix_delta_H_SC} is replaced by
\begin{align}\label{eq:appendix_delta_H_QM}
    \delta H = -\frac{1}{2} \left\{ \hat{\mathcal{P}}_{\alpha}, E_{\alpha}(\hat{\mathbf{r}}) \right\}.
\end{align}
We note that this reduces to $\hat{\mathbf{P}}_{\text{AC}}=\gamma \hbar \hat{\mathbf{v}} \times \hat{\mathbf{z}}/c^{2}$, which is the AC polarization operator introduced in Ref.~\cite{Tang26PRL}.

We choose the electric field in the following form
\begin{align}
    E_{\alpha}(\mathbf{r}) &= \frac{p}{q} \text{sin}(qr_{\beta}).
\end{align}
Then, we have
\begin{align}
    \delta \mathcal{E} &= - Q_{\alpha \beta} \partial_{\beta} E_{\alpha} = - p \, Q_{\alpha \beta} \, \text{cos}(qr_{\beta}).
\end{align}
Accordingly, the component of the local zero-point density conjugate to the electric-field gradient is
\begin{align}
    \delta \mathcal{E}_{\text{ZP}}(\mathbf{r}) &= -p \, Q_{\alpha \beta}^{\text{ZP}} \, \text{cos}(qr_{\beta}),
\end{align}
where $Q_{\alpha \beta}^{\text{ZP}}$ is the zero-point fluctuation-induced AC-EQ. From this, we obtain
\begin{align}\label{eq:appendix_Q_form}
    Q_{\alpha \beta}^{\text{ZP}} &= - \lim\limits_{q \to 0} \lim\limits_{p \to 0} \frac{2}{pV} \int d\mathbf{r} \, \delta \mathcal{E}_{\text{ZP}} (\mathbf{r}) \, \text{cos}(qr_{\beta}).
\end{align}

Now, we consider the contribution of $\mathcal{E}_{\text{ZP}}^{(2)}$. Reference~\cite{Tang26PRL} explicitly showed that this zero-point contribution does not generate the magnon OAM for a fixed magnetic moment (i.e., $\bm{\mu}_{n\mathbf{k}} = \mu\hat{\mathbf{z}}$). Here, we generalize it to an arbitrary electric-field-gradient component and to the momentum-dependent magnetic moment considered in the present work. To first order in the perturbation, the local energy variation $\mathcal{E}_{\text{ZP}}^{(2)}$ contains
\begin{align}\label{eq:appendix_ZP2_reduced}
    \delta \mathcal{E}_{\text{ZP}}^{(2)}(\mathbf{r}) = \frac{1}{2}\sum_{n>0, \mathbf{k}} \biggr[ &\braket{\psi_{n\mathbf{k}}|H_{0}|\delta \psi_{n\mathbf{k}}} + \braket{\psi_{n\mathbf{k}}|\delta H| \psi_{n\mathbf{k}}} \notag
    \\[2mm]
    &+ \braket{\delta \psi_{n\mathbf{k}}|H_{0}| \psi_{n\mathbf{k}}} \biggr],
\end{align}
where $\ket{\psi} = e^{i\mathbf{k} \cdot \mathbf{r}}\ket{u_{n\mathbf{k}}}$. The second term is proportional to the applied field $E_{\alpha} \propto \text{sin}(qr_{\beta})$ and has no projection onto the $\text{cos}(qr_{\beta})$ component defining the response of $\partial_{\beta}E_{\alpha}$. Thus, we have
\begin{align}
    \delta \mathcal{E}_{\text{ZP}}^{(2)}(\mathbf{r}) &= \frac{1}{2} \sum_{n>0, \mathbf{k}} \left[ \braket{\psi_{n\mathbf{k}}|H_{0}|\delta \psi_{n\mathbf{k}}} + (\text{c.c.})\right].
\end{align}
The perturbed state is given by
\begin{align}\label{eq:appendix_perturbed_state}
    \ket{\delta u_{n\mathbf{k}}} = -\frac{p}{2iq} \sum_{m=1}^{2N} \biggr[& \frac{\ket{u_{m,\mathbf{k} + \mathbf{q}}}\braket{\tilde{u}_{m,\mathbf{k}+\mathbf{q}}|\bar{\mathcal{P}}_{\alpha}(\mathbf{k} + \mathbf{q}, \mathbf{k})|u_{n,\mathbf{k}}}}{\bar{\mathcal{E}}_{n\mathbf{k}} - \bar{\mathcal{E}}_{m,\mathbf{k} + \mathbf{q}}}\notag
    \\[2mm]
    &- (\mathbf{q} \rightarrow -\mathbf{q}) \biggr].
\end{align}
By combining Eqs.~\eqref{eq:appendix_Q_form}, \eqref{eq:appendix_ZP2_reduced} and \eqref{eq:appendix_perturbed_state}, we finally obtain
\begin{align}\label{eq:appendix_Qzp2_final}
    Q_{\alpha \beta}^{\text{ZP}(2)} \propto \lim\limits_{q \rightarrow 0}\sum_{n > 0, m \mathbf{k}}\frac{1}{2iq}\biggr[& \braket{\tilde{u}_{m,\mathbf{k}+\mathbf{q}}|\bar{\mathcal{P}}_{\alpha}(\mathbf{k} + \mathbf{q} , \mathbf{k})|u_{n\mathbf{k}}}\notag
    \\[2mm]
    &\times \braket{\tilde{u}_{n\mathbf{k}}|u_{m,\mathbf{k} + \mathbf{q}}} - (\text{c.c.})\biggr],
\end{align}
where $\bar{\mathcal{P}}_{\alpha}(\mathbf{k}+\mathbf{q}, \mathbf{k}) = [\bar{\mathcal{P}}_{\alpha}(\mathbf{k} + \mathbf{q}) + \bar{\mathcal{P}}_{\alpha}(\mathbf{k})]/2$. Here, $\bar{\mathcal{E}}_{n\mathbf{k}}$ denotes the $n$th diagonal element of $\bar{\mathcal{E}}_{\mathbf{k}} = \sigma_{3}\mathcal{E}_{\mathbf{k}}$ [Eq.~\eqref{eq:appendix_bar_E}], $\bar{\mathcal{P}}_{\alpha} = \sigma_{3}\mathcal{P}_{\alpha}$, and $\bra{\tilde{u}_{n\mathbf{k}}} = s_{n}\bra{u^{\text{L}}_{n\mathbf{k}}}$. Equation~\eqref{eq:appendix_Qzp2_final} is the generalization of the zero-point expression presented in Ref.~\cite{Tang26PRL}.

We divide the summation into $m = n$ and $m \neq n$ cases. In the case \uppercase\expandafter{\romannumeral1}, where $m = n$ (diagonal), we have
\begin{widetext}
\begin{align}\label{eq:appendix_ZP2_case1}
    &\lim\limits_{q \rightarrow 0}\frac{1}{q} \left[ \braket{\tilde{u}_{n,\mathbf{k}+\mathbf{q}}|\bar{\mathcal{P}}_{\alpha}(\mathbf{k} + \mathbf{q} , \mathbf{k})|u_{n\mathbf{k}}} \braket{\tilde{u}_{n\mathbf{k}}|u_{n,\mathbf{k} + \mathbf{q}}} - (\text{c.c.})\right] = \braket{\partial_{k_{\beta}}\tilde{u}_{n\mathbf{k}}|\bar{\mathcal{P}}_{\alpha}|u_{n\mathbf{k}}} + \braket{\tilde{u}_{n\mathbf{k}}|\bar{\mathcal{P}}_{\alpha}|u_{n\mathbf{k}}} \braket{\tilde{u}_{n\mathbf{k}}|\partial_{k_{\beta}}u_{n\mathbf{k}}} - (\text{c.c.}) \notag
    \\[2mm]
    &= \sum_{l \neq n} \left[ \braket{\partial_{k_{\beta}}\tilde{u}_{n\mathbf{k}}|u_{l\mathbf{k}}} \braket{\tilde{u}_{l\mathbf{k}}|\bar{\mathcal{P}}_{\alpha}|u_{n\mathbf{k}}} - (\text{c.c.}) \right] .
\end{align}
In the case \uppercase\expandafter{\romannumeral2}, where $m \neq n$ (off-diagonal), we obtain
\begin{align}\label{eq:appendix_ZP2_case2}
    &\lim\limits_{q \rightarrow 0}\sum_{m \neq n}\frac{1}{q} \left[\braket{\tilde{u}_{m,\mathbf{k}+\mathbf{q}}|\bar{\mathcal{P}}_{\alpha}(\mathbf{k} + \mathbf{q} , \mathbf{k})|u_{n\mathbf{k}}} \braket{\tilde{u}_{n\mathbf{k}}|u_{m,\mathbf{k} + \mathbf{q}}} - (\text{c.c.}) \right] = \sum_{m \neq n}\left[ \braket{\tilde{u}_{m\mathbf{k}}|\bar{\mathcal{P}}_{\alpha}|u_{n\mathbf{k}}} \braket{\tilde{u}_{n\mathbf{k}}|\partial_{k_{\beta}}u_{m\mathbf{k}}} - (\text{c.c.}) \right] \notag
    \\[2mm]
    &= -\sum_{m \neq n} \left[ \braket{\partial_{k_{\beta}}\tilde{u}_{n\mathbf{k}}|u_{m\mathbf{k}}} \braket{\tilde{u}_{m\mathbf{k}}|\bar{\mathcal{P}}_{\alpha}|u_{n\mathbf{k}}} - (\text{c.c.})\right] .
\end{align}
\end{widetext}
Therefore, the diagonal and off-diagonal contributions cancel exactly. We note that the $\mathcal{O}(q)$ corrections arising from the expansion $\bar{\mathcal{P}}_{\alpha}(\mathbf{k} + \mathbf{q}, \mathbf{k}) = \bar{\mathcal{P}}_{\alpha}(\mathbf{k}) + q\partial_{k_{\beta}}\bar{\mathcal{P}}_{\alpha}/2 + \mathcal{O}(q^{2})$ vanish both in Eqs.~\eqref{eq:appendix_ZP2_case1} and \eqref{eq:appendix_ZP2_case2}.

Finally, we consider the contribution of $\mathcal{E}_{\text{ZP}}^{(1)}$. Unlike $\mathcal{E}^{\text{ZP}(2)}$, which is a spectral quantity, $\mathcal{E}^{\text{ZP}(1)}$ is the trace of the BdG Hamiltonian itself. Thus, its first-order response is determined entirely by the explicit variation of the Hamiltonian. Thus, it is sufficient to examine whether this variation contains a term linear in the electric-field gradient. For a finite-$\mathbf{q}$ electric field, we have
\begin{align}
    \delta \mathcal{E}^{(1)}_{\text{ZP}}(\mathbf{q}) &= -\frac{1}{4} \int_{\mathbf{k}} \text{tr} [\delta H_{\mathbf{k}+\frac{\mathbf{q}}{2}, \mathbf{k}-\frac{\mathbf{q}}{2}}].
\end{align}
From Eq.~\eqref{eq:appendix_delta_H_QM}, we have
\begin{align}
    \delta H_{\mathbf{k}+\frac{\mathbf{q}}{2}, \mathbf{k}-\frac{\mathbf{q}}{2}} &= -\frac{E_{\alpha}(\mathbf{q})}{2}\left[ \mathcal{P}_{\alpha}\left(\mathbf{k}+\frac{\mathbf{q}}{2}\right) + \mathcal{P}_{\alpha}\left(\mathbf{k}-\frac{\mathbf{q}}{2}\right)\right].
\end{align}
The expression in brackets is even in $\mathbf{q}$ and therefore contains no term linear in $q_{\beta}$. It implies that
\begin{align}
    \delta \mathcal{E}^{(1)}_{\text{ZP}}(\mathbf{q}) &= \frac{E_{\alpha}(\mathbf{q})}{4} \int_{\mathbf{k}} \text{tr} \left[ \mathcal{P}_{\alpha}\left(\mathbf{k}\right)\right] + \mathcal{O}(q^{2})
\end{align}
also contains no contribution proportional to $\partial_{\beta}E_{\alpha}$. Therefore, $\mathcal{E}_{\text{ZP}}^{(1)}$ generates no contribution conjugate to $\partial_{\beta}E_{\alpha}$ and
\begin{align}
    Q_{\alpha \beta}^{\text{ZP}(1)} &= - \lim\limits_{q \to 0} \lim\limits_{p \to 0} \frac{2}{pV} \int d\mathbf{r} \, \delta \mathcal{E}_{\text{ZP}}^{(1)} (\mathbf{r}) \, \text{cos}(qr_{\beta}) = 0.
\end{align}

Combining these two results, we conclude that the harmonic zero-point fluctuation energy does not contribute to the AC-EQ
\begin{align}
    Q_{\alpha \beta}^{\text{ZP}} &= Q_{\alpha \beta}^{\text{ZP}(1)} + Q_{\alpha \beta}^{\text{ZP}(2)} = 0.
\end{align}

\bibliography{paper}

@misc{suppl_ref,
  note = {See {S}upplemental {M}aterial for further details, which includes Refs.~\cite{Riedl16PRb,Neumann20PRL,stock10PRb}.},
  year = {}
}

@article{Xiao10RMP,
  title = {Berry phase effects on electronic properties},
  author = {Xiao, Di and Chang, Ming-Che and Niu, Qian},
  journal = {Rev. Mod. Phys.},
  volume = {82},
  issue = {3},
  pages = {1959--2007},
  numpages = {0},
  year = {2010},
  month = {Jul},
  publisher = {American Physical Society},
  doi = {10.1103/RevModPhys.82.1959},
  url = {https://link.aps.org/doi/10.1103/RevModPhys.82.1959}
}

@article{Aharonov84PRL,
  title = {Topological Quantum Effects for Neutral Particles},
  author = {Aharonov, Y. and Casher, A.},
  journal = {Phys. Rev. Lett.},
  volume = {53},
  issue = {4},
  pages = {319--321},
  numpages = {0},
  year = {1984},
  month = {Jul},
  publisher = {American Physical Society},
  doi = {10.1103/PhysRevLett.53.319},
  url = {https://link.aps.org/doi/10.1103/PhysRevLett.53.319}
}

@article{Cimmino89PRL,
  title = {Observation of the topological {A}haronov-{C}asher phase shift by neutron interferometry},
  author = {Cimmino, A. and Opat, G. I. and Klein, A. G. and Kaiser, H. and Werner, S. A. and Arif, M. and Clothier, R.},
  journal = {Phys. Rev. Lett.},
  volume = {63},
  issue = {4},
  pages = {380--383},
  numpages = {0},
  year = {1989},
  month = {Jul},
  publisher = {American Physical Society},
  doi = {10.1103/PhysRevLett.63.380},
  url = {https://link.aps.org/doi/10.1103/PhysRevLett.63.380}
}

@article{Resta98PRL,
  title = {Quantum-Mechanical Position Operator in Extended Systems},
  author = {Resta, Raffaele},
  journal = {Phys. Rev. Lett.},
  volume = {80},
  issue = {9},
  pages = {1800--1803},
  numpages = {0},
  year = {1998},
  month = {Mar},
  publisher = {American Physical Society},
  doi = {10.1103/PhysRevLett.80.1800},
  url = {https://link.aps.org/doi/10.1103/PhysRevLett.80.1800}
}

@article{Xiao05PRL,
  title = {Berry Phase Correction to Electron Density of States in Solids},
  author = {Xiao, Di and Shi, Junren and Niu, Qian},
  journal = {Phys. Rev. Lett.},
  volume = {95},
  issue = {13},
  pages = {137204},
  numpages = {4},
  year = {2005},
  month = {Sep},
  publisher = {American Physical Society},
  doi = {10.1103/PhysRevLett.95.137204},
  url = {https://link.aps.org/doi/10.1103/PhysRevLett.95.137204}
}

@article{Thonhauser05PRL,
  title = {Orbital Magnetization in Periodic Insulators},
  author = {Thonhauser, T. and Ceresoli, Davide and Vanderbilt, David and Resta, R.},
  journal = {Phys. Rev. Lett.},
  volume = {95},
  issue = {13},
  pages = {137205},
  numpages = {4},
  year = {2005},
  month = {Sep},
  publisher = {American Physical Society},
  doi = {10.1103/PhysRevLett.95.137205},
  url = {https://link.aps.org/doi/10.1103/PhysRevLett.95.137205}
}

@article{Bernevig05PRL,
  title = {Orbitronics: The Intrinsic Orbital Current in $p$-Doped Silicon},
  author = {Bernevig, B. Andrei and Hughes, Taylor L. and Zhang, Shou-Cheng},
  journal = {Phys. Rev. Lett.},
  volume = {95},
  issue = {6},
  pages = {066601},
  numpages = {4},
  year = {2005},
  month = {Aug},
  publisher = {American Physical Society},
  doi = {10.1103/PhysRevLett.95.066601},
  url = {https://link.aps.org/doi/10.1103/PhysRevLett.95.066601}
}

@article{Shi07PRL,
  title = {Quantum Theory of Orbital Magnetization and Its Generalization to Interacting Systems},
  author = {Shi, Junren and Vignale, G. and Xiao, Di and Niu, Qian},
  journal = {Phys. Rev. Lett.},
  volume = {99},
  issue = {19},
  pages = {197202},
  numpages = {4},
  year = {2007},
  month = {Nov},
  publisher = {American Physical Society},
  doi = {10.1103/PhysRevLett.99.197202},
  url = {https://link.aps.org/doi/10.1103/PhysRevLett.99.197202}
}

@article{Kontani09PRL,
  title = {Giant Orbital {H}all Effect in Transition Metals: Origin of Large Spin and Anomalous {H}all Effects},
  author = {Kontani, H. and Tanaka, T. and Hirashima, D. S. and Yamada, K. and Inoue, J.},
  journal = {Phys. Rev. Lett.},
  volume = {102},
  issue = {1},
  pages = {016601},
  numpages = {4},
  year = {2009},
  month = {Jan},
  publisher = {American Physical Society},
  doi = {10.1103/PhysRevLett.102.016601},
  url = {https://link.aps.org/doi/10.1103/PhysRevLett.102.016601}
}

@article{Mena14PRL,
  title={Spin-wave spectrum of the quantum ferromagnet on the pyrochlore lattice {L}u$_{2}${V}$_{2}${O}$_{7}$},
  author={Mena, M and Perry, RS and Perring, TG and Le, MD and Guerrero, S and Storni, M and Adroja, DT and R{\"u}egg, Ch and McMorrow, DF},
  journal={Phys. Rev. Lett.},
  volume={113},
  number={4},
  pages={047202},
  year={2014},
  publisher={APS},
  doi = {10.1103/PhysRevLett.113.047202},
  url = {https://doi.org/10.1103/PhysRevLett.113.047202}
}

@article{Go18PRL,
  title = {Intrinsic Spin and Orbital {H}all Effects from Orbital Texture},
  author = {Go, Dongwook and Jo, Daegeun and Kim, Changyoung and Lee, Hyun-Woo},
  journal = {Phys. Rev. Lett.},
  volume = {121},
  issue = {8},
  pages = {086602},
  numpages = {6},
  year = {2018},
  month = {Aug},
  publisher = {American Physical Society},
  doi = {10.1103/PhysRevLett.121.086602},
  url = {https://link.aps.org/doi/10.1103/PhysRevLett.121.086602}
}

@article{Neumann20PRL,
  title = {Orbital Magnetic Moment of Magnons},
  author = {Neumann, Robin R. and Mook, Alexander and Henk, J\"urgen and Mertig, Ingrid},
  journal = {Phys. Rev. Lett.},
  volume = {125},
  issue = {11},
  pages = {117209},
  numpages = {7},
  year = {2020},
  month = {Sep},
  publisher = {American Physical Society},
  doi = {10.1103/PhysRevLett.125.117209},
  url = {https://link.aps.org/doi/10.1103/PhysRevLett.125.117209}
}

@article{Fishman22PRL,
  title = {Orbital Angular Momentum of Magnons in Collinear Magnets},
  author = {Fishman, Randy S. and Gardner, Jason S. and Okamoto, Satoshi},
  journal = {Phys. Rev. Lett.},
  volume = {129},
  issue = {16},
  pages = {167202},
  numpages = {7},
  year = {2022},
  month = {Oct},
  publisher = {American Physical Society},
  doi = {10.1103/PhysRevLett.129.167202},
  url = {https://link.aps.org/doi/10.1103/PhysRevLett.129.167202}
}

@article{Gobel24PRL,
  title = {Orbital {H}all Effect Accompanying Quantum {H}all Effect: {L}andau Levels Cause Orbital Polarized Edge Currents},
  author = {G\"obel, B\"orge and Mertig, Ingrid},
  journal = {Phys. Rev. Lett.},
  volume = {133},
  issue = {14},
  pages = {146301},
  numpages = {7},
  year = {2024},
  month = {Oct},
  publisher = {American Physical Society},
  doi = {10.1103/PhysRevLett.133.146301},
  url = {https://link.aps.org/doi/10.1103/PhysRevLett.133.146301}
}

@article{Liu25PRL,
  title = {Quantum Correction to the Orbital {H}all Effect},
  author = {Liu, Hong and Cullen, James H. and Arovas, Daniel P. and Culcer, Dimitrie},
  journal = {Phys. Rev. Lett.},
  volume = {134},
  issue = {3},
  pages = {036304},
  numpages = {7},
  year = {2025},
  month = {Jan},
  publisher = {American Physical Society},
  doi = {10.1103/PhysRevLett.134.036304},
  url = {https://link.aps.org/doi/10.1103/PhysRevLett.134.036304}
}

@article{Tang26PRL,
  title = {Proper Theory of Magnon Orbital Angular Momentum at Equilibrium},
  author = {Tang, Junyu and Cheng, Ran},
  journal = {Phys. Rev. Lett.},
  volume = {136},
  issue = {23},
  pages = {236702},
  numpages = {7},
  year = {2026},
  month = {Jun},
  publisher = {American Physical Society},
  doi = {10.1103/fgt6-bhmq},
  url = {https://link.aps.org/doi/10.1103/fgt6-bhmq}
}

@article{Liu2011PRL,
  title={Electric control of spin currents and spin-wave logic},
  author={Liu, Tianyu and Vignale, G},
  journal={Phys. Rev. Lett.},
  volume={106},
  number={24},
  pages={247203},
  year={2011},
  publisher={APS},
  doi = {10.1103/PhysRevLett.106.247203},
  url = {https://doi.org/10.1103/PhysRevLett.106.247203}
}

@article{Zhang2014PRL,
  title={Electric-field coupling to spin waves in a centrosymmetric ferrite},
  author={Zhang, Xufeng and Liu, Tianyu and Flatt{\'e}, Michael E and Tang, Hong X},
  journal={Phys. Rev. Lett.},
  volume={113},
  number={3},
  pages={037202},
  year={2014},
  publisher={APS},
  doi = {10.1103/PhysRevLett.113.037202},
  url = {https://doi.org/10.1103/PhysRevLett.113.037202}
}

@article{Sundaram99PRb,
  title = {Wave-packet dynamics in slowly perturbed crystals: Gradient corrections and {B}erry-phase effects},
  author = {Sundaram, Ganesh and Niu, Qian},
  journal = {Phys. Rev. B},
  volume = {59},
  issue = {23},
  pages = {14915--14925},
  numpages = {0},
  year = {1999},
  month = {Jun},
  publisher = {American Physical Society},
  doi = {10.1103/PhysRevB.59.14915},
  url = {https://link.aps.org/doi/10.1103/PhysRevB.59.14915}
}

@article{Ceresoli06PRb,
  title = {Orbital magnetization in crystalline solids: Multi-band insulators, {C}hern insulators, and metals},
  author = {Ceresoli, Davide and Thonhauser, T. and Vanderbilt, David and Resta, R.},
  journal = {Phys. Rev. B},
  volume = {74},
  issue = {2},
  pages = {024408},
  numpages = {13},
  year = {2006},
  month = {Jul},
  publisher = {American Physical Society},
  doi = {10.1103/PhysRevB.74.024408},
  url = {https://link.aps.org/doi/10.1103/PhysRevB.74.024408}
}

@article{Ceresoli10PRb,
  title = {First-principles theory of orbital magnetization},
  author = {Ceresoli, Davide and Gerstmann, Uwe and Seitsonen, Ari P. and Mauri, Francesco},
  journal = {Phys. Rev. B},
  volume = {81},
  issue = {6},
  pages = {060409(R)},
  numpages = {4},
  year = {2010},
  month = {Feb},
  publisher = {American Physical Society},
  doi = {10.1103/PhysRevB.81.060409},
  url = {https://link.aps.org/doi/10.1103/PhysRevB.81.060409}
}

@article{Lopez12PRb,
  title = {Wannier-based calculation of the orbital magnetization in crystals},
  author = {Lopez, M. G. and Vanderbilt, David and Thonhauser, T. and Souza, Ivo},
  journal = {Phys. Rev. B},
  volume = {85},
  issue = {1},
  pages = {014435},
  numpages = {12},
  year = {2012},
  month = {Jan},
  publisher = {American Physical Society},
  doi = {10.1103/PhysRevB.85.014435},
  url = {https://link.aps.org/doi/10.1103/PhysRevB.85.014435}
}

@article{Shindou13PRb,
  title = {Topological chiral magnonic edge mode in a magnonic crystal},
  author = {Shindou, Ryuichi and Matsumoto, Ryo and Murakami, Shuichi and Ohe, Jun-ichiro},
  journal = {Phys. Rev. B},
  volume = {87},
  issue = {17},
  pages = {174427},
  numpages = {11},
  year = {2013},
  month = {May},
  publisher = {American Physical Society},
  doi = {10.1103/PhysRevB.87.174427},
  url = {https://link.aps.org/doi/10.1103/PhysRevB.87.174427}
}

@article{Cheng13PRb,
  title = {Microscopic derivation of spin-transfer torque in ferromagnets},
  author = {Cheng, Ran and Niu, Qian},
  journal = {Phys. Rev. B},
  volume = {88},
  issue = {2},
  pages = {024422},
  numpages = {6},
  year = {2013},
  month = {Jul},
  publisher = {American Physical Society},
  doi = {10.1103/PhysRevB.88.024422},
  url = {https://link.aps.org/doi/10.1103/PhysRevB.88.024422}
}

@article{Hanke16PRb,
  title = {Role of {B}erry phase theory for describing orbital magnetism: From magnetic heterostructures to topological orbital ferromagnets},
  author = {Hanke, J.-P. and Freimuth, F. and Nandy, A. K. and Zhang, H. and Bl\"ugel, S. and Mokrousov, Y.},
  journal = {Phys. Rev. B},
  volume = {94},
  issue = {12},
  pages = {121114(R)},
  numpages = {5},
  year = {2016},
  month = {Sep},
  publisher = {American Physical Society},
  doi = {10.1103/PhysRevB.94.121114},
  url = {https://link.aps.org/doi/10.1103/PhysRevB.94.121114}
}

@article{Nakata17PRb,
  title = {Magnonic topological insulators in antiferromagnets},
  author = {Nakata, Kouki and Kim, Se Kwon and Klinovaja, Jelena and Loss, Daniel},
  journal = {Phys. Rev. B},
  volume = {96},
  issue = {22},
  pages = {224414},
  numpages = {14},
  year = {2017},
  month = {Dec},
  publisher = {American Physical Society},
  doi = {10.1103/PhysRevB.96.224414},
  url = {https://link.aps.org/doi/10.1103/PhysRevB.96.224414}
}

@article{Cheon18PRb,
  title={Nonreciprocal spin waves in a chiral antiferromagnet without the {D}zyaloshinskii-{M}oriya interaction},
  author={Cheon, Suik and Lee, Hyun-Woo and Cheong, Sang-Wook},
  journal={Phys. Rev. B},
  volume={98},
  number={18},
  pages={184405},
  year={2018},
  publisher={APS},
  doi = {10.1103/PhysRevB.98.184405},
  url = {https://doi.org/10.1103/PhysRevB.98.184405}
}

@article{Liu19PRb,
  title = {Magnon quantum anomalies in {W}eyl ferromagnets},
  author = {Liu, Tianyu and Shi, Zheng},
  journal = {Phys. Rev. B},
  volume = {99},
  issue = {21},
  pages = {214413},
  numpages = {24},
  year = {2019},
  month = {Jun},
  publisher = {American Physical Society},
  doi = {10.1103/PhysRevB.99.214413},
  url = {https://link.aps.org/doi/10.1103/PhysRevB.99.214413}
}

@article{Pezo22PRB,
  title = {Orbital {H}all effect in crystals: Interatomic versus intra-atomic contributions},
  author = {Pezo, Armando and Garc\'{\i}a Ovalle, Diego and Manchon, Aur\'elien},
  journal = {Phys. Rev. B},
  volume = {106},
  issue = {10},
  pages = {104414},
  numpages = {6},
  year = {2022},
  month = {Sep},
  publisher = {American Physical Society},
  doi = {10.1103/PhysRevB.106.104414},
  url = {https://link.aps.org/doi/10.1103/PhysRevB.106.104414}
}

@article{Fishman23PRb,
  title = {Gauge-invariant measure of the magnon orbital angular momentum},
  author = {Fishman, Randy S.},
  journal = {Phys. Rev. B},
  volume = {107},
  issue = {21},
  pages = {214434},
  numpages = {8},
  year = {2023},
  month = {Jun},
  publisher = {American Physical Society},
  doi = {10.1103/PhysRevB.107.214434},
  url = {https://link.aps.org/doi/10.1103/PhysRevB.107.214434}
}

@article{Lee24PRb,
  title = {Orbital {E}delstein effect of electronic itinerant orbital motion at edges},
  author = {Lee, Jongjun M. and Park, Min Ju and Lee, Hyun-Woo},
  journal = {Phys. Rev. B},
  volume = {110},
  issue = {13},
  pages = {134436},
  numpages = {8},
  year = {2024},
  month = {Oct},
  publisher = {American Physical Society},
  doi = {10.1103/PhysRevB.110.134436},
  url = {https://link.aps.org/doi/10.1103/PhysRevB.110.134436}
}

@article{Lee25PRb,
  title = {Universal intrinsic orbital dynamics from {B}erry curvature in electronic two-band systems},
  author = {Lee, Jongjun M.},
  journal = {Phys. Rev. B},
  volume = {112},
  issue = {5},
  pages = {054441},
  numpages = {8},
  year = {2025},
  month = {Aug},
  publisher = {American Physical Society},
  doi = {10.1103/42mj-sh8x},
  url = {https://link.aps.org/doi/10.1103/42mj-sh8x}
}

@article{Kuzmenko25PRb,
  title = {The {A}haronov-{C}asher phase is geometrical and not topological},
  author = {Kuzmenko, Igor and Band, Y. B. and Avishai, Yshai},
  journal = {Phys. Rev. B},
  volume = {112},
  issue = {2},
  pages = {L020501},
  numpages = {5},
  year = {2025},
  month = {Jul},
  publisher = {American Physical Society},
  doi = {10.1103/rm8w-8rmm},
  url = {https://link.aps.org/doi/10.1103/rm8w-8rmm}
}

@article{Lee26PRb,
  title = {Anatomy of the modern theory of orbital magnetism from first principles: Term-by-term analysis in the gauge-covariant formalism},
  author = {Lee, Hojun and Baek, Insu and Sastges, Mirco and Mokrousov, Yuriy and Lee, Hyun-Woo and Go, Dongwook},
  journal = {Phys. Rev. B},
  volume = {113},
  issue = {21},
  pages = {214449},
  numpages = {32},
  year = {2026},
  month = {Jun},
  publisher = {American Physical Society},
  doi = {10.1103/k1l2-g57n},
  url = {https://link.aps.org/doi/10.1103/k1l2-g57n}
}

@article{Yamamoto26PRb,
  title = {Ferroelectricity in a magnon {B}ose-{E}instein condensate: Nonreciprocal superfluidity, exceptional points, and {M}ajorana bosons},
  author = {Yamamoto, Kazuki and Kawakami, Takuto and Koshino, Mikito},
  journal = {Phys. Rev. B},
  volume = {114},
  issue = {3},
  pages = {034501},
  numpages = {12},
  year = {2026},
  month = {Jul},
  publisher = {American Physical Society},
  doi = {10.1103/b6nh-6cj6},
  url = {https://link.aps.org/doi/10.1103/b6nh-6cj6}
}

@article{Bahramy11PRb,
  title = {Origin of giant bulk {R}ashba splitting: Application to {B}i{T}e{I}},
  author = {Bahramy, M. S. and Arita, R. and Nagaosa, N.},
  journal = {Phys. Rev. B},
  volume = {84},
  issue = {4},
  pages = {041202(R)},
  numpages = {4},
  year = {2011},
  month = {Jul},
  publisher = {American Physical Society},
  doi = {10.1103/PhysRevB.84.041202},
  url = {https://link.aps.org/doi/10.1103/PhysRevB.84.041202}
}

@article{Boliasova2025PRb,
  title = {Magnonic {A}haronov-{C}asher effect and electric field control of chirality-dependent spin-wave dynamics in antiferromagnets},
  author = {Boliasova, O. O. and Krivoruchko, V. N.},
  journal = {Phys. Rev. B},
  volume = {111},
  issue = {17},
  pages = {174440},
  numpages = {10},
  year = {2025},
  month = {May},
  publisher = {American Physical Society},
  doi = {10.1103/PhysRevB.111.174440},
  url = {https://link.aps.org/doi/10.1103/PhysRevB.111.174440}
}

@article{Boyer97PRa,
  title = {Proposed {A}haronov-{C}asher effect: Another example of an {A}haronov-{B}ohm effect arising from a classical lag},
  author = {Boyer, Timothy H.},
  journal = {Phys. Rev. A},
  volume = {36},
  issue = {10},
  pages = {5083--5086},
  numpages = {0},
  year = {1987},
  month = {Nov},
  publisher = {American Physical Society},
  doi = {10.1103/PhysRevA.36.5083},
  url = {https://link.aps.org/doi/10.1103/PhysRevA.36.5083}
}

@article{Bakke09PRa,
  title = {Relativistic {L}andau quantization for a neutral particle},
  author = {Bakke, K. and Furtado, C.},
  journal = {Phys. Rev. A},
  volume = {80},
  issue = {3},
  pages = {032106},
  numpages = {7},
  year = {2009},
  month = {Sep},
  publisher = {American Physical Society},
  doi = {10.1103/PhysRevA.80.032106},
  url = {https://link.aps.org/doi/10.1103/PhysRevA.80.032106}
}

@article{Busch23PRR,
  title = {Orbital {H}all effect and orbital edge states caused by $s$ electrons},
  author = {Busch, Oliver and Mertig, Ingrid and G\"obel, B\"orge},
  journal = {Phys. Rev. Res.},
  volume = {5},
  issue = {4},
  pages = {043052},
  numpages = {9},
  year = {2023},
  month = {Oct},
  publisher = {American Physical Society},
  doi = {10.1103/PhysRevResearch.5.043052},
  url = {https://link.aps.org/doi/10.1103/PhysRevResearch.5.043052}
}

@article{Choi23NAT,
  title = "{Observation of the orbital {H}all effect in a light metal {T}i}",
  author = {Choi, Young-Gwan and Jo, Daegeun and Ko, Kyung-Hun and Go, Dongwook and Kim, Kyung-Han and Park, Hee Gyum and Kim, Changyoung and Min, Byoung-Chul and Choi, Gyung-Min and Lee, Hyun-Woo},
  journal = {Nature},
  volume = {619},
  issue = {7968},
  pages = {52-56},
  year = {2023},
  month = {Jul},
  doi = {10.1038/s41586-023-06101-9},
  url = {https://doi.org/10.1038/s41586-023-06101-9}
}

@article{Yamamoto25CP,
  title = {Electromagnetic response in dipole superfluids},
  author = {Yamamoto, Kazuki and Kawakami, Takuto and Koshino, Mikito},
  journal = {Commun. Phys.},
  volume = {8},
  issue = {1},
  pages = {171},
  year = {2025},
  doi = {10.1038/s42005-025-02088-z},
  url = {https://doi.org/10.1038/s42005-025-02088-z}
}

@article{Jo24NPJS,
  title = {Spintronics meets orbitronics: Emergence of orbital angular momentum in solids},
  author = {Jo, Daegeun and Go, Dongwook and Choi, Gyung-Min and Lee, Hyun-Woo},
  journal = {npj Spintronics},
  volume = {2},
  issue = {1},
  pages = {19},
  year = {2024},
  doi = {10.1038/s44306-024-00023-6},
  url = {https://doi.org/10.1038/s44306-024-00023-6}
}

@article{Onose10SCI,
  title={Observation of the magnon {H}all effect},
  author={Onose, Y and Ideue, T and Katsura, H and Shiomi, Y and Nagaosa, N and Tokura, Y},
  journal={Science},
  volume={329},
  number={5989},
  pages={297--299},
  year={2010},
  publisher={American Association for the Advancement of Science},
  doi = {10.1126/science.1188260},
  url = {https://doi.org/10.1126/science.1188260}
}

@article{Mignani91JPa,
  title = {Aharonov-{C}asher effect and geometrical phases},
  author = {R Mignani},
  journal = {J. Phys. A: Math. Gen.},
  year = {1991},
  month = {Apr},
  publisher = {IOP Publishing},
  volume = {24},
  number = {8},
  pages = {L421},
  doi = {10.1088/0305-4470/24/8/006},
  url = {https://doi.org/10.1088/0305-4470/24/8/006}
}

@article{Fishman23JPc,
  title = {Exact results for the orbital angular momentum of magnons on honeycomb lattices},
  author = {Fishman, Randy S and Lindsay, Lucas and Okamoto, Satoshi},
  journal = {J. Phys. Condens. Matter},
  year = {2022},
  month = {Nov},
  publisher = {IOP Publishing},
  volume = {35},
  number = {1},
  pages = {015801},
  doi = {10.1088/1361-648X/ac9a28},
  url = {https://doi.org/10.1088/1361-648X/ac9a28}
}

@article{Go24NanoLett,
  title = {Magnon Orbital {N}ernst Effect in Honeycomb Antiferromagnets without Spin–Orbit Coupling},
  author = {Go, Gyungchoon and An, Daehyeon and Lee, Hyun-Woo and Kim, Se Kwon},
  journal = {Nano Lett.},
  volume = {24},
  issue = {20},
  pages = {5968-5974},
  year = {2024},
  month = {May},
  doi = {10.1021/acs.nanolett.4c00430},
  url = {https://doi.org/10.1021/acs.nanolett.4c00430}
}

@article{Wang25AElecM,
  title = {Orbitronics: Mechanisms, Materials and Devices},
  author = {Wang, Ping and Chen, Feng and Yang, Yuhe and Hu, Shuai and Li, Yue and Wang, Wenhong and Zhang, Delin and Jiang, Yong},
  journal = {Adv. Electron. Mater.},
  volume = {11},
  number = {5},
  pages = {2400554},
  doi = {10.1002/aelm.202400554},
  url = {https://doi.org/10.1002/aelm.202400554},
  year = {2025}
}

@article{Colpa78Physa,
  title = {Diagonalization of the quadratic boson hamiltonian},
  author = {J.H.P. Colpa},
  journal = {Physica A},
  volume = {93},
  number = {3},
  pages = {327-353},
  year = {1978},
  issn = {0378-4371},
  doi = {10.1016/0378-4371(78)90160-7},
  url = {https://doi.org/10.1016/0378-4371(78)90160-7}
}

@article{Basso20EPL,
  title = "{Electric field effect on spin waves: Role of magnetic moment current}",
  author = {Basso, Vittorio and Ansalone, Patrizio},
  journal = {EPL},
  volume = {130},
  number = {1},
  pages = {17008},
  year = {2020},
  month = {May},
  publisher = {EDP Sciences, IOP Publishing and Società Italiana di Fisica},
  doi = {10.1209/0295-5075/130/17008},
  url = {https://doi.org/10.1209/0295-5075/130/17008}
}

@article{Go21EPL,
  title = "{Orbitronics: Orbital currents in solids}",
  author = {Dongwook Go and Daegeun Jo and Hyun-Woo Lee and Mathias Kläui and Yuriy Mokrousov},
  journal = {EPL},
  volume = {135},
  number = {3},
  pages = {37001},
  year = {2021},
  month = {Sep},
  publisher = {EDP Sciences, IOP Publishing and Società Italiana di Fisica},
  doi = {10.1209/0295-5075/ac2653},
  url = {https://dx.doi.org/10.1209/0295-5075/ac2653}
}

@article{Bakke19EPJP,
  title="{A semiclassical treatment of the interaction of non-uniform electric fields with the electric quadrupole moment of a neutral particle}",
  author = {Bakke, K.},
  journal = {Eur. Phys. J. Plus},
  volume = {134},
  pages = {76},
  year = {2019},
  month = {Feb},
  doi = {10.1140/epjp/i2019-12489-1},
  url = {https://doi.org/10.1140/epjp/i2019-12489-1}
}

@article{Stone16At,
  title={Table of nuclear electric quadrupole moments},
  author={Stone, N. J.},
  journal={At. Data Nucl. Data Tables},
  volume={111-112},
  pages={1-28},
  year={2016},
  publisher={Elsevier},
  url = {https://doi.org/10.1016/j.adt.2015.12.002},
  doi = {10.1016/j.adt.2015.12.002}
}

@article{Cysne26arXiv,
  title = {Orbital {H}all effect from orbital magnetic moments of {B}loch states: The role of a new correction term},
  author = {Cysne, Tarik P. and Souza, Ivo and Rappoport, Tatiana G.},
  journal = {Phys. Rev. Res.},
  volume = {8},
  issue = {3},
  pages = {033086},
  numpages = {12},
  year = {2026},
  month = {Jul},
  publisher = {American Physical Society},
  doi = {10.1103/xflp-2y1c},
  url = {https://link.aps.org/doi/10.1103/xflp-2y1c}
}

@misc{Sastges26arXiv,
  title={Modern approach to orbital {H}all effect based on {W}annier picture of solids}, 
  author={Mirco Sastges and Insu Baek and Hojun Lee and Hyun-Woo Lee and Yuriy Mokrousov and Dongwook Go},
  year={2026},
  eprint={2604.08280},
  archivePrefix={arXiv},
  primaryClass={cond-mat.mes-hall},
  url={https://doi.org/10.48550/arXiv.2604.08280}
}

@article{Jeon26arXiv,
  title = {Comparing the orbital angular momentum and magnetic moment of magnons in the kagome antiferromagnet with negative spin chirality},
  author = {Jeon, Youngjae and Lee, Jongjun M. and Cheon, Suik and Lee, Hyun-Woo},
  journal = {Phys. Rev. B},
  pages = {},
  year = {2026},
  month = {Aug},
  publisher = {American Physical Society},
  doi = {10.1103/ddpk-l8jc},
  url = {https://link.aps.org/doi/10.1103/ddpk-l8jc}
}

@misc{Lee2026arXiv,
  title={Observation of current-induced orbital quadrupole accumulation},
  author={Lee, Geun-Hee and Ji, Yubin and Park, Yongho and An, Changmin and Ko, San and Ko, Hye-Won and Lim, Jinseob and Oh, Jung Hyun and Mahfouzi, Farzad and Park, Byong-Guk and others},
  year={2026},
  eprint={2608.22838},
  archivePrefix={arXiv},
  primaryClass={cond-mat.mtrl-sci},
  url={https://doi.org/10.48550/arXiv.2608.22838}
}

@article{Yamasaki2025STAM,
  title={Sum rules for {X}-ray circular and linear dichroism based on complete magnetic multipole basis},
  author={Yamasaki, Y and Ishii, Y and Sasabe, N},
  journal={Sci. Technol. Adv. Mater.},
  volume={26},
  number={1},
  pages={2513217},
  year={2025},
  publisher={Taylor \& Francis},
  doi = {10.1080/14686996.2025.2513217},
  url = {https://doi.org/10.1080/14686996.2025.2513217}
}
\end{document}